\documentclass[num-refs]{wiley-article}
\usepackage{color}
\usepackage{ulem}

\usepackage{siunitx}
\usepackage{hhline}
\usepackage{multicol}
\usepackage{url,booktabs,multirow}
\usepackage{caption}
\usepackage{subcaption}
\usepackage{natbib}
\usepackage{color}
\usepackage{enumitem}
\usepackage{lineno,hyperref,amsmath}
\usepackage{algorithm2e}
\usepackage{comment}

\RestyleAlgo{ruled}
\SetKwComment{Comment}{/* }{ */}

\papertype{Original Article}
\title{Releasing survey microdata with exact cluster locations and additional privacy safeguards}

\author[1\authfn{1}]{Till Koebe}
\author[1\authfn{1}]{Alejandra Arias-Salazar}
\contrib[\authfn{1}]{Equally contributing authors.}

\affil[1]{Department of Economics, Freie Universität Berlin, Berlin, Germany}
\corraddress{Till Koebe, Department of Economics, Freie Universität Berlin, Garystrasse 21, 14195 Berlin, Germany}
\corremail{till.koebe@fu-berlin.de}

\fundinginfo{The authors did not receive any specific funding for this work.}

\runningauthor{Koebe and Arias-Salazar}

\begin{document}

\maketitle

%\linenumbers

\begin{abstract}

Household survey programs around the world publish fine-granular georeferenced microdata to support research on the interdependence of human livelihoods and their surrounding environment. To safeguard the respondents' privacy, micro-level survey data is usually (pseudo)-anonymized through deletion or perturbation procedures such as obfuscating the true location of data collection. This, however, poses a challenge to emerging approaches that augment survey data with auxiliary information on a local level. Here, we propose an alternative microdata dissemination strategy that leverages the utility of the original microdata with additional privacy safeguards through synthetically generated data using generative models.  We back our proposal with experiments using data from the 2011 Costa Rican census and satellite-derived auxiliary information. Our strategy reduces the respondents' re-identification risk for any number of disclosed attributes by 60-80\% even under re-identification attempts.

% Please include a maximum of seven keywords
 %\keywords{Census updating, auxiliary information, SPREE, satellite imagery, survey data}
\end{abstract}

\section{Introduction}

% In recent years, natural disasters and extreme weather events once again turned a spotlight on the interdependencies of human livelihoods and their surrounding environment. Likewise, household survey programs around the world increasingly started to publish fine-granular georeferenced microdata to make investigating those links possible.  
Since almost hundred years, sample surveys are dominating knowledge generation in empirical research. The advantages of survey sampling are obvious: with an appropriate sampling design representative results for a population can be collected by surveying only a fraction of it. With computer assistance, the time from collecting data to publishing results can be sped up significantly \citep{granello2004}. 
Two trends, however, increasingly challenge the way data is collected via surveys. On the one hand, the growing demand for fast and granular information drives up sample size and thus costs. As a response, recent years have seen a large amount of academic research on augmenting surveys with secondary data from non-traditional data sources such as social networks, mobile phones or remote sensing in order to overcome shortcomings in coverage, frequency and granularity with applications in fields as diverse as population dynamics \citep{Stevens2015DisaggregatingData, leasure2020national}, socio-demographic analysis \citep{Pokhriyal2017, Schmid2017ConstructingSenegalb, subash2018satellite, fatehkia2020relative, Chi2022}, policy targeting \citep{Blumenstock2018, Aiken2022}, environmental mapping \citep{Grace2019} and health research \citep{Brown2014, arambepola2020spatiotemporal}. 
This augmentation is usually done via geographic matching, i.e. combining area-level averages \citep{Koebe2020BetterModelling}. Since the number of matched areas corresponds to the sample size for subsequent supervised learning tasks, finding the smallest common geographical denominator is essential to avoid running into small sample problems. However, this is not always trivial as sample surveys usually provide data only for a fraction of small geographic areas.
%, which, in contrast to most non-traditional data sources, follow administrative delineations, but satellite imagery and mobile phone metadata are usually provided as pixels and as point locations of the corresponding network antennas, respectively. 
On the other hand, digital transformations across various sectors such as health care have led to an explosion of digital personal data. % [REFERENCE]. 
It is the abundance of secondary data that amplifies re-identification risks in published surveys as some of the information could be used to link pseudoanonymized survey responses back to the actual respondents \citep{armstrong1999geographically, Kroll2016, West2017}. % [MORE REFERENCES NEEDED].
Together with new privacy regulations such as the European General Data Protection Regulation (GDPR) %\citep{GDPR2018}
%this has motivated the search for data processing and storing methods that reconcile data utility with data protection since.%data considered sensitive in a sense that the interviewee does not want pieces of personal information be shared publicly. 
%Consequently, 
this calls for additional precautionary measures to safeguard the individual's privacy. For aggregated data releases, the introduction of differential privacy has provided a solid mathematical framework to manage re-identification risks independent of a potential attacker's capabilities or prior knowledge \citep{Dwork2008}. %When it comes to publishing survey microdata, 
With regard to microdata dissemination strategies, a common de-identification practice today is a combination of deletion and perturbation procedures, which include removing (unique) identifiers such as first and last name and replacing the individual's true location with aggregated (i.e. area-level) and randomized information (see e.g. \cite{AndresGeo-Indistinguishability:Systems, templ2017SDC, sdcspatial2019}). %Furthermore, as mentioned in \cite{templ2017} additionally to synthetic data generation techniques, there are two other kind of methods: perturbative and non-perturbartive techniques. These methods are characterized by applying procedures (with different levels of invasion) to the pre-defined key and sensitive variables in the original data, producing a modified data file that is intended to maintain its usefulness in terms of statistical analysis and reducing at the same time, the risk of re-identify units. However, for statistical data that should be released periodically (e.g. National Statistical Surveys), these techniques carry some limitations: 1. specifying what information should be protected is not trivial, specially when many variables are considered sensitive, 2. definitions of sensitive variables may change over time, 3. the knowledge, capabilities and external sources that an intruder could use to identify units can also vary from one year to another. 

For example, in the Demographic and Health Survey (DHS), a major global household survey program, urban survey clusters are re-located within a 2km-radius and rural clusters within a 5km-, sometimes even 10km-radius \citep{DHS2013}. This location privacy procedure has two main advantages: it does not affect the quality of the remaining (non-spatial) survey information and it reduces the need for other privacy safeguards, e.g. deleting or perturbing sensitive information. However, it does not provide a similar rigorous measure for privacy protection as already small sets of attributes can quickly increase the chances of re-identification, even in incomplete, pseudonymous datasets \citep{Rocher2019EstimatingModels}. In addition, it obviously affects the utility of the published data when it comes to matching with auxiliary data as this type of analysis relies on the congruence of its geographic links \citep{elkies2015scrambling, warren2016influence, Blankespoor2021, hunter2021working}.

In that regard, advances in synthetic data generation have introduced new ways to narrow the void between information loss and privacy protection. These methods allow for the generation of synthetic records that resemble the real data by reproducing relationships learned from the latter. %\cite{Templ2017simPop} categorizes synthetic data generation methods into three groups: combinatorial optimization, synthetic reconstruction, and model-based data generation. 
While all approaches have in common that they try to capture the joint distribution in the original data, the ways to do so vastly differ. For example, \cite{drechsler2008new} and \cite{heldal2019synthetic} use imputation processes to decompose the multidimensional joint distribution into conditional univariate distributions. \cite{alfons2011simulation} and \cite{Templ2017simPop} use parametric models in combination with conditional re-sampling to synthesize hierarchical relationships. As an alternative to these fully parametric approaches, \cite{reiter2005} and \cite{wang2012multiple} make use of classification and regression trees (CART), while more recently, \cite{Li2014, zhang2017, Rocher2019EstimatingModels, Sun2019, torkzadehmahani2019, Xu2019} and others have used Bayesian networks, Generative Adversarial Networks or copulas to capture the underlying linear and non-linear relationships between the attributes.
%These approaches have been largely proposed to find a balance between providing quality microdata and preserving privacy.
%, and have been successful under particular conditions.

The challenge for data producers is %to provide options to users that meet their needs, 
to define adequate microdata dissemination strategies that allow users to satisfy their needs, i.e. release survey microdata that can be used for statistical analysis and that are compatible with other sources of information allowing to answer new and more detailed research questions and -- at the same time -- it must be ensured that the identities of the respondents are protected. 
%Most microdata dissemination strategies rely on a combination of licensing requirements and deletion or perturbation procedures. For National Statistical Offices and international programs such as DHS and IPUMS, the challenge is to provide useful micro data, allowing users to keep contributing with statistical analysis and research.
In that regard, the Spatial Data Repository of the DHS program \citep{dhswebsite} is a good example for facilitating new types of research by combining survey microdata with geospatial covariates and gridded interpolation surfaces. However, also those products are based on perturbed cluster locations, thus incurring a certain information loss.

In this paper, we propose an alternative microdata dissemination strategy that leverages the utility of the original microdata with additional privacy safeguards through copula-generated synthetic data. Specifically, we propose to adopt a strategy of publishing two sets of micro-level survey data: first, the original microdata stripped of geographic identifiers below the strata-level. Second, synthetic microdata including the true cluster locations. We show in an experiment using Costa Rican census data from 2011 and satellite-derived auxiliary information from WorldPop \citep{WorldPopwww.worldpop.org-SchoolofGeographyandEnvironmentalScienceUniversityofSouthamptonDepartmentofGeographyandGeosciencesUniversityofLouisvilleDepartementdeGeographie2018WorldPopb1} that we can reduce the re-identification risk vis-à-vis common spatial perturbation procedures, while maintaining data utility for non-spatial analysis and improving data utility for spatial analysis.
%we investigate the capability of two methods to maximize the utility of microdata and reduce the risk of identifying individuals and their sensitive information, namely: geomasking (spatial anonymization) and synthetic data generation via copulas. Among the data generation methods, we focus on a copula-based approach due to practical and theoretical convenience: for up-coming years, only new nationally representative margins are required to updated the synthetic microdata file, the methodology implemented is transparent, and well-documented tools are available to users with important features such as data transformation and constraints specification (see: https://sdv.dev). 
% In this sense, we conduct an experiment based on real data to compare the performance of geomasked survey data, and synthetic survey data, interacting with spatial data. We seek to answer the following questions in a holistic way: can we improve the efficiency of the estimates for small areas? i.e. increase the utility of the data for statistical analysis, for example, to provide quality information in small areas? and,  is it possible to reduce the risk of re-identifying the survey respondents when the two sources of information (survey and spatial data) are combined? %maybe add:
% Finally, we aim to provide recommendations to data producers on potential dissemination strategies based on the scope of these methods.

From the plethora of options, we choose copulas as our synthetic data generation approach. Copulas facilitate fine-tuning as they allow us to model the marginal distributions separately from the joint distribution. Dating back to 1959 \citep{sklar1959} with diverse applications since, their theoretical properties are well understood. In comparison with alternatives like GANs, copula-based synthetic data generation has lower computational cost \citep{Sun2019} and it is easier to interpret \citep{kamthe2021copula}. Furthermore, the procedure is in general less cumbersome, in comparison with the steps followed by \cite{alfons2011simulation} to generate the synthetic population data AAT-SILC (\textit{Artificial Austrian Statistics on Income and Living Conditions} \citep{AMELI2011}. Finally, copulas are also attractive for data producers such as National Statistical Offices as only new nationally representative margins are required to update the synthetic microdata file (cf. \cite{Koebe2022}). In addition, well-documented open-source tools such as the Synthetic Data Vault \citep{sdv2022} are available to users with important features such as data transformation and constraints specification.

\section{Results}

\subsection{Geomasking to obfuscate true survey locations}

We consider a survey $D$ as a random sample with sample size $n_D$ from a given population $C$. Our unit of observations are individuals $i$ living together in a household $h$. Each individual is described by a set of attributes denoted as $X_1, X_2, \dots X_d$. While different sampling designs are possible, we assume a commonly used complex design for larger household surveys such as the DHS: a stratified two-stage cluster design. In the first stage, the primary sampling units (PSUs) - usually enumeration areas from the latest census denoted as $j$ - are selected for each stratum $s$ with a probability proportional to (population) size $\pi_j$. In the second stage, households within each selected PSU are sampled with a fixed probability $\pi_{h|j}$. Consequently, the sampling weights defined as the inverses of the household-level inclusion probabilities are given for each stratum separately by:

\begin{equation}
w_{hj} = \frac{1}{\pi_{hj}},  \hspace{1cm} \pi_{hj}  = \pi_{h|j} * \pi_j
\quad \text{with} \quad \pi_j = \frac{n_s}{N_s}.
\end{equation}

PSUs, called \textit{clusters} in the following, are geo-located as point locations $r_j$ via their geographic centroids. In the following, we describe the original survey attributes together with the original geo-locations of the clusters as our \textit{true} survey. The true survey builds our starting point for further anonymization approaches, notably the geomasking approach and the copula-based synthetic data generation approach. We follow the geomasking methodology outlined in \cite{DHS2013} by perturbing the centroids of the selected clusters within a given larger administrative area $l$ using a rejection sampling procedure described in Algorithm \ref{alg:one}:

\normalem %%%% disable auto underline

\begin{algorithm}
\caption{Geomasked survey: DHS cluster displacement algorithm}\label{alg:one}
\For{$j \in D$}{
    \While{$r_j^{\text{masked}} \notin l_{r_j}$}{
        $\text{angle} \gets U_{[0, 360]} * \frac{\pi}{180}$ \Comment*[r]{Random displacement angle}
        \If{$j$ is Urban}{
        $\text{dist} \gets U_{[0,2000]}$ \Comment*[r]{Random displacement distance (in meters) for urban clusters}
        }
        \If{$j$ is Rural}{
            \eIf{$j$ is selected as 1\% of rural clusters}{
            $\text{dist} \gets U_{[0,10000]}$ \Comment*[r]{Random displacement distance for 1\% of rural clusters}
            }{
            $\text{dist} \gets U_{[0,5000]}$ 
            }
        }
        $r_{x, j}^{\text{masked}} \gets r_{x, j} + \text{dist} * \cos(\text{angle})$ \Comment*[r]{Displaced x-coordinate}
        $r_{y, j}^{\text{masked}} \gets r_{y, j} + \text{dist} * \sin(\text{angle})$ \Comment*[r]{Displaced y-coordinate}
    }
}

\end{algorithm}

\ULforem %%%% enable auto underline

We denote the masked point locations of the selected clusters with the superscript \textit{masked}. As the overall inclusion probability for a household is not affected by geomasking, direct estimates and corresponding variances for area-level aggregates $l$ and above remain the same. However, this does not hold for area-level aggregates smaller than $l$. In the following, we describe the original survey attributes together with the masked locations of the clusters as our \textit{geomasked} survey.

For our experiment using Costa Rican census data from 2011, point locations for the corresponding enumeration areas are not available. Therefore, we randomly sample them from the smallest available area denoted with $k$ - in our experiment the districts (at the same time the zip codes) in Costa Rica. The zip code therefore corresponds to the smallest geographic identifier available in the survey. Through the displacement procedure, roughly 30\% of the clusters are assigned to a new zip code, representing approx. 30\% of the sampled individuals in each simulation round.

\subsection{Copula-based synthetic data generation} \label{section:results-copula}

As an alternative to geomasking, we use synthetically generated survey attributes for protecting the respondents' privacy while keeping the true point locations of the selected clusters. To do so, we fit a Gaussian copula model on the original survey attributes $X_d$ and sample from the learned joint distribution for each cluster individually with the originally sample size $n_j$. A copula allows to describe the dependence structure - also called \textit{association structure} - independently from the marginal distributions (also called \textit{allocation structure}). Several copula families are available. We focus on the Gaussian copula that allows us to represent the association structure of random variables irrespective of their true distribution through a multivariate standard normal distribution \cite{patki2016synthetic}. Since we also assume the marginals to be normally distributed, which may certainly constitute a mis-specification for some of the variables, we regard the results rather as a lower bound in terms of goodness-of-fit. Further, a copula is uniquely defined only for continuous variables \cite{jeong2016copula}, meaning that in principle, copulas cannot model non-continuous variables. Since socio-economic surveys are largely made up of categorical variables, data transformation, e.g. via one-hot encoding, is needed. In addition, we impose constraints on the marginals to account for censoring (e.g. to avoid negative synthetic age records) or between-variable dependencies (e.g. female and male household members need to add up to the total household size) via rejection sampling.

Thus, the process to generate synthetic data from a survey dataset $\tilde{D}$ with transformed categorical attributes (details of the data transformation are described in Algorithms \ref{alg:three} and \ref{alg:four} in Section \ref{section:methods}) using a Gaussian copula model is summarized in Algorithm \ref{alg:two}: 

\normalem %%%% disable auto underline

\begin{algorithm}
\caption{Synthetic survey: Copula-based synthetic data generation algorithm}\label{alg:two}
\textbf{Input $\tilde{D} = (\tilde{X}_1, \dots, \tilde{X}_d)$}\\
\textbf{Output $\tilde{S} =  (\tilde{Y}_1, \dots, \tilde{Y}_d)$}

\BlankLine

\For{$s \in \tilde{D}$}{
%  $X_d \gets$ univariate vector with $X_d \sim \mathcal{N}(\mu, \sigma^2)$  \\
%\eIf{$\tilde{X}_a|\tilde{X}_b$}{
%$\tilde{D} = (\tilde{X}_a,\tilde{X}_b), \hspace{2pt} a \cup b, \hspace{2pt} a= (\tilde{X}_1,\dots , \tilde{X}_m) \hspace{1pt}, \hspace{1pt} b = (\tilde{X}_{m+1}, \dots, \tilde{X}_w)$\\
%$\Psi \gets$ Estimated marginal distribution of $\tilde{X}_a$ conditional to $\tilde{X}_b$ with $\psi_m \sim \mathcal{N}(\mu^*, \sigma^{2*} )$ \\
%$\Sigma \gets$ Estimated correlation matrix of $\Psi=\Psi_{a|b}$\\
%$U \gets F_{a|b}(\Psi_{a|b}) $ \\
%&\Comment*[r]{Probability integral transforms}
%$C^G_{\Sigma} (u_1, \dots, u_m) \gets \phi_{\Sigma_{a|b}} \big(\phi_{1}^{-1}(u_1|\tilde{x}_b), \dots , \phi_{m}^{-1}(u_m|\tilde{x}_b)\big) $ %\Comment*[r]{m-dimensional Gaussian copula
%$C^G_{\Sigma_{a|b}} (u_1, \dots, u_m|\tilde{x}_b) \gets \phi_{\Sigma_{a|b}} \big(\phi_{1}^{-1}(u_1|\tilde{x}_b), \dots , \phi_{m}^{-1}(u_m|\tilde{x}_b)\big) $ \Comment*[r]{m-dimensional Gaussian copula
%}
%}{
$\Psi \gets$ Estimated marginal distributions of $\tilde{X}$ with $\psi_d \sim \mathcal{N}(\mu, \sigma^2)$ \\
$\Sigma \gets$ Estimated covariance matrix of $\Psi$ \\
$U \gets F(\Psi) $  \Comment*[r]{Probability integral transforms}
$C^G_{\Sigma} (u_1, \dots, u_d) \gets \phi_{\Sigma} \big(\phi_{1}^{-1}(u_1), \dots , \phi_{d}^{-1}(u_d)\big)$ \Comment*[r]{d-dimensional Gaussian copula}

\For{$j \in \tilde{D}_s$}{
\For{$i \gets 1 $  \textbf{\text{to}} $n_{D_j}$ }{
\While{$\textbf{y}^{\{i\}}$ not meets \textit{constraints}}{ 
% \For{$m \gets 1 $  \textbf{\text{to}} $\tilde{X}_{m_j}$ }{
% Draw $z_m \sim \mathcal{N}(0,1)$
% }
% $L \gets  \text{Cholesky}(\Sigma)$ \Comment*[r]{Cholesky decomposition of $\Sigma$}

% $\textbf{v} \gets L\textbf{z}$ \Comment*[r]{Correlation properties for unrelated $z_d$}

$\textbf{v} \sim \mathcal{N}(\mu,\Sigma)$

$\tilde{\textbf{y}}^{\{i\}} \gets F^{-1}\big(\phi_1(v_1), \dots, \phi_d(v_d) \big)$  \Comment*[r]{Convert back to original space}
}
}
}
}
\end{algorithm}

\ULforem %%%% enable auto underline

$\phi_{\Sigma}$ is the cumulative distribution function (cdf) of a multivariate normal distribution with $\mathcal{N}(\mu,\Sigma)$ and $\phi_d$ the cdf of a standard normal distribution. By fitting our model to the true survey, it learns the parameters of both the allocation and association structure, i.e. of the marginal distributions \textbf{$\Psi$} and the multivariate Gaussian copula $C_{\Sigma}^G(u_1,\dots,u_d)$. Based on these learned relationships, new synthetic records $\tilde{\textbf{y}}^{\{i\}}$ are sampled from the multivariate probability function $c_{\Sigma}^G(\textbf{u})$ using the inverse probability integral transform for each component $F_d^{-1}(u_d)$ (cf. \cite{janke2021}). Since we sample in our experiment for each cluster individually to ensure a synthetic cluster-level sample size of exactly $n_j$, we use the parameters of a conditional multivariate normal distribution. In case no conditions are applied, the scenario is simplified to drawing from a multivariate standard normal distribution. In the following, we call the synthetic attributes together with the true cluster locations our \textit{synthetic} survey. Further details about the copula-based synthetic data generation procedure can be found in the Section \ref{section:methods} and in \cite{nelsen2007introduction}.

\begin{figure}[ht!]
\captionsetup[subfigure]{justification=centering}
     \centering
     \begin{subfigure}[c]{0.3\linewidth}
         \centering
         \includegraphics[width=\textwidth]{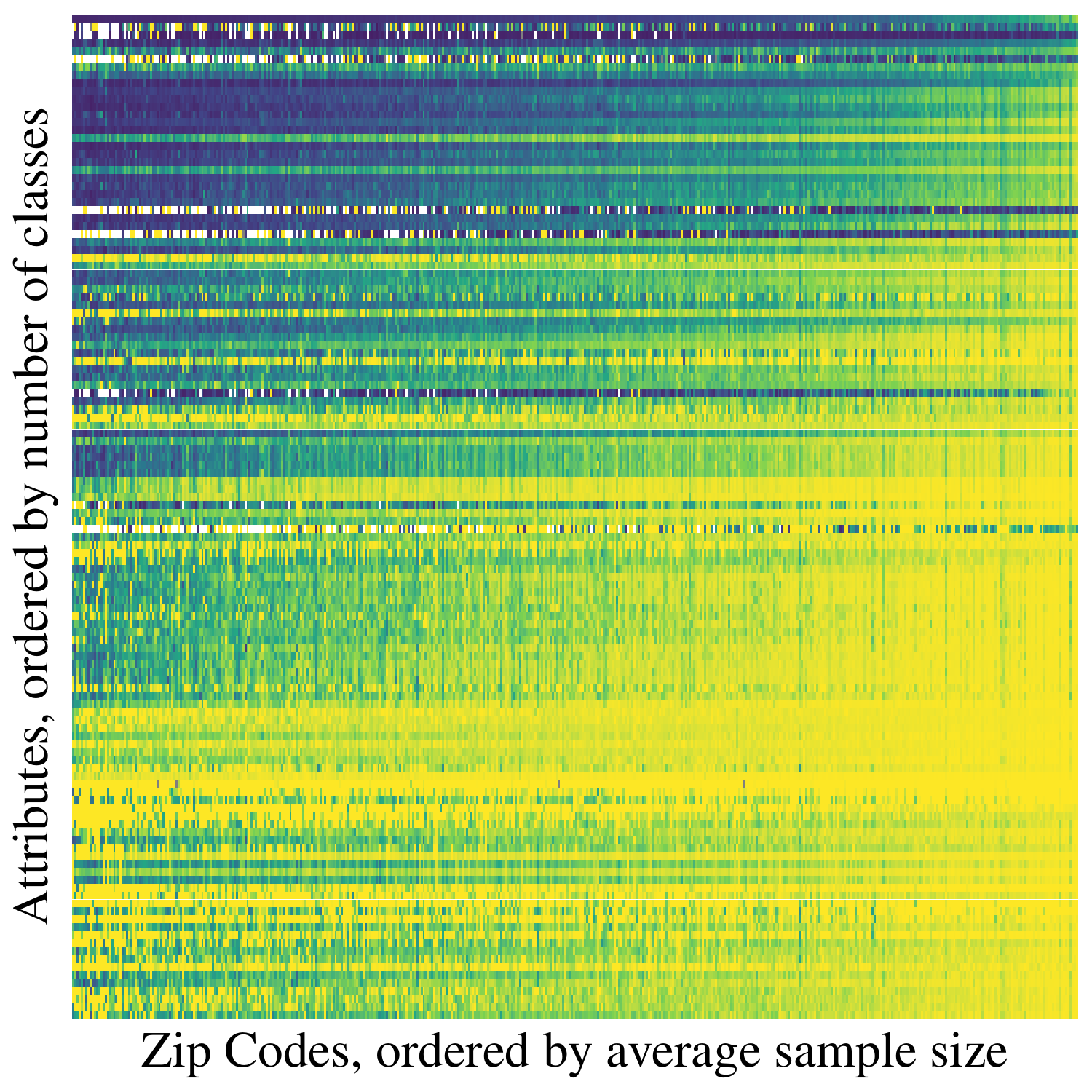}
         \caption{True survey}
         \label{fig:kl-true}
     \end{subfigure}
     \begin{subfigure}[c]{0.3\linewidth}
         \centering
         \includegraphics[width=\textwidth]{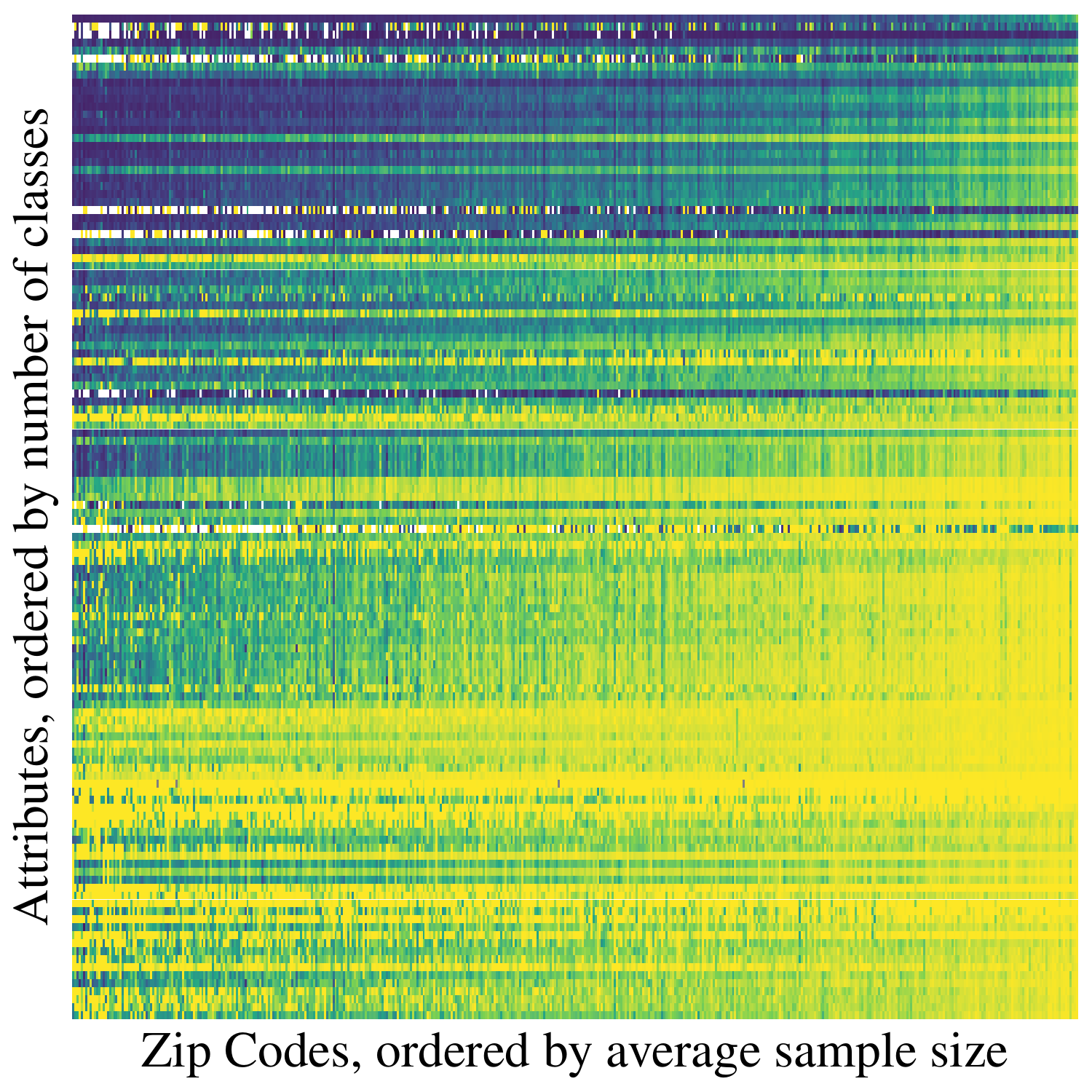}
         \caption{Geomasked survey}
         \label{fig:kl-geo}
     \end{subfigure}
     \begin{subfigure}[c]{0.3\linewidth}
         \centering
         \includegraphics[width=\textwidth]{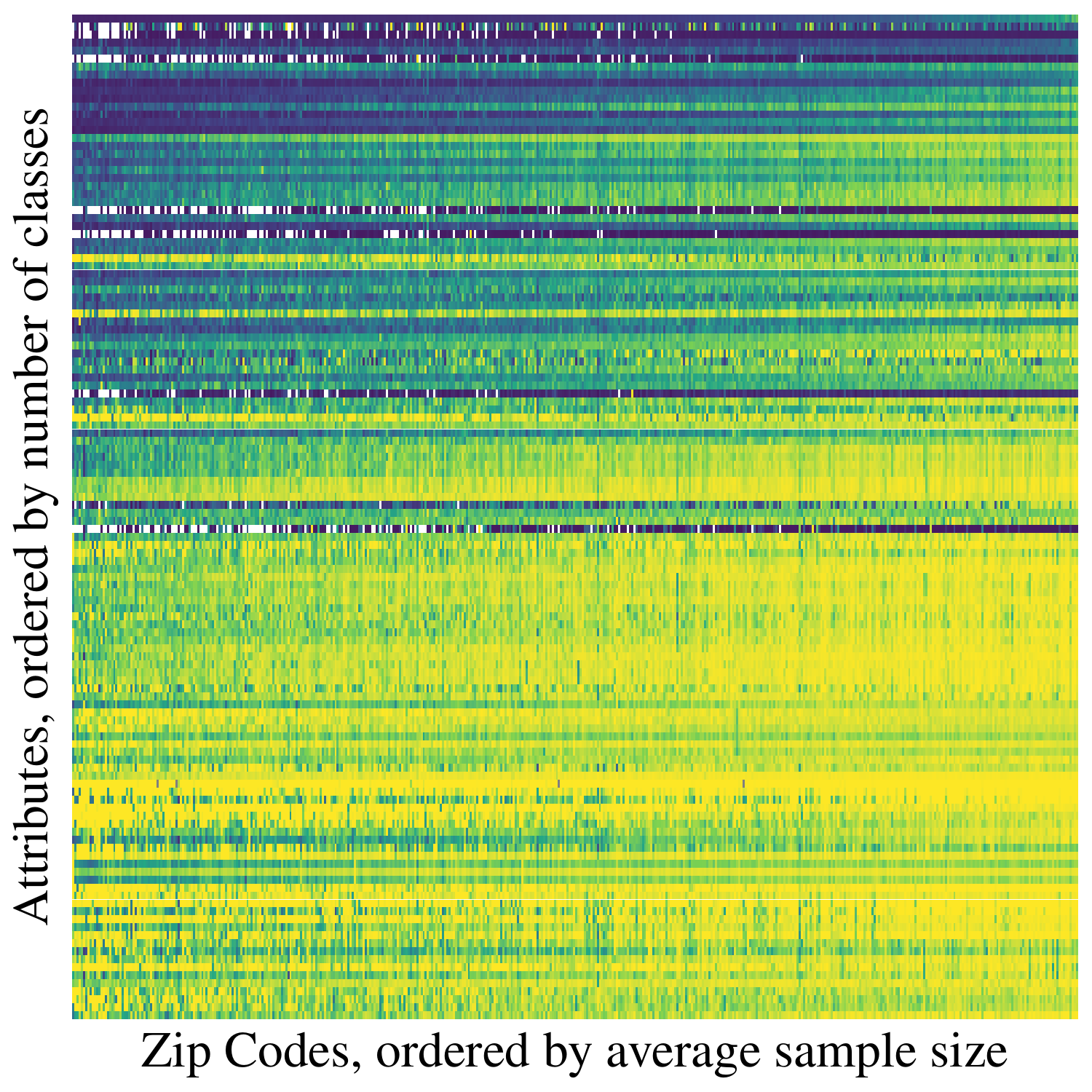}
         \caption{Synthetic survey}
         \label{fig:kl-syn}
     \end{subfigure}
     \begin{subfigure}[c]{0.07\linewidth}
         \centering
         \includegraphics[width=\textwidth]{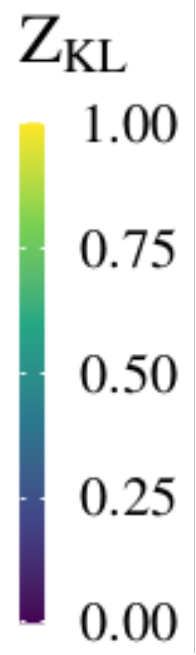}
        %  \caption{Synthetic survey}
         \label{fig:legend}
     \end{subfigure}
        \caption{\textbf{Normalized Kullback-Leibler divergence (in bits) from the true census distribution for each attribute and zip code, averaged across 100 simulation rounds.}\\ The attributes on the y-axis are ordered by their respective number of classes, the zip codes on the x-axis are ordered by their average sample size across simulation rounds. Values close to one (yellow) represent little divergence from the true census distribution and therefore indicate a high goodness-of-fit. The number of attribute classes range from two to 111. Across attributes and zip codes, the true survey scores best with $Z_{KL} = 0.76$ in total, followed by the synthetic survey with $Z_{KL} = 0.74$ and the geomasked survey at $Z_{KL} = 0.73$.}
        \label{fig:kl}
\end{figure}

Figure \ref{fig:kl} provides a first impression on the overall goodness-of-fit of the three different survey datasets. Specifically, Figures \ref{fig:kl-true} - \ref{fig:kl-syn} show the normalized KL divergence $Z_{KL}$ of the survey attributes $\textbf{y}$ from the true census attributes $\textbf{x} \in C$ defined as

\begin{equation}
    Z_{KL}(F_{d,k}(X_{d,k})||F_{d,k}(Y_{d,k})) = \frac{1}{1 + D_{KL}(F_{d,k}(X_{d,k})||F_{d,k}(Y_{d,k}))}
\end{equation}

averaged across simulation runs for each attribute $d$ and zip code $k$, respectively. 

Clearly visible is a gradient from the top left to the bottom right indicating that the overall goodness-of-fit of the sample distributions improve the larger the underlying sample sizes and the lower the number of classes per categorical attribute. In addition, as expected, attributes with high levels of non-response (visible through the white spots across the horizontal axis) are stronger affected by sampling and anonymization compared to attributes with little or no non-response.

\subsection{Population uniqueness of survey respondents}

To approach the utility-risk trade-off in (pseudo)-anonymized microdata, we define two risk-related measures: the population uniqueness of the survey respondents and the re-identification risk of a sensitive attribute in the original data using the perturbed data. We define population uniqueness $\Xi_\textbf{x}$ as the share of survey respondents being unique in the population $C$ for a given set of attributes $\textbf{x} = (X_1,\dots,X_d)$:

\begin{equation}
    \Xi_\textbf{x} = \frac{1}{n_D}\sum_i^{n_D}\mathbb{1}_{\textbf{x}^{\{i\}}} \quad \text{with} \quad \mathbb{1}_{\textbf{x}^{\{i\}}} = 
    \begin{cases}
      1, & \text{if}\ \textbf{x}^{\{i\}} \text{ unique in } C(\textbf{x}) \\
      0, & \text{otherwise}
    \end{cases}
\end{equation}

% \begin{equation*}
%     \Xi_\textbf{x} = \frac{1}{n_D}\sum_i^{n_D}[\textbf{x}^{\{i\}} \text{unique in} C(\textbf{x})]
% \end{equation*}

\begin{figure}[ht!]
    \centering
    \includegraphics[width=\textwidth]{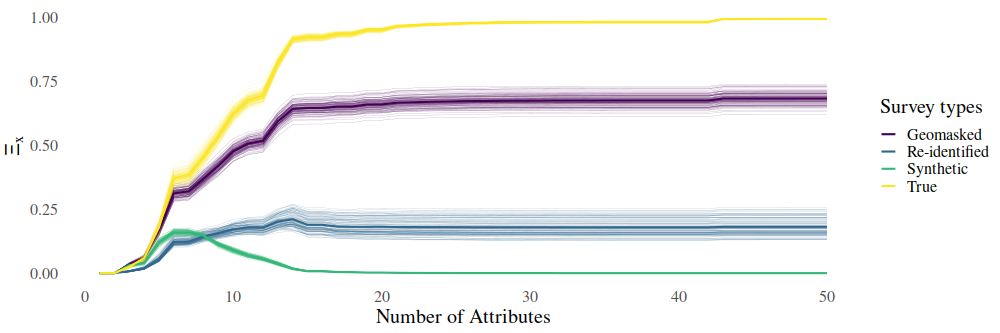}
    \caption{\textbf{Population uniqueness across survey types.}\\
    Share of population-unique survey respondents for 100 simulation runs. The thick lines represent the average population uniqueness across the 100 simulation runs, the thin lines individual simulation runs. In the \textit{true} survey, no attribute is perturbed. In the \textit{geomasked} survey, the cluster identifier is perturbed. In the \textit{synthetic} survey, all variables but the cluster identifier are perturbed. In the \textit{re-identified} survey, the synthetic survey is used to predict the "private" attribute -- i.e. the cluster -- in the original dataset along the lines of the proposed microdata dissemination strategy. The re-identified original survey is then used to calculate population uniqueness. Both the re-identified and the synthetic survey provide significant privacy gains vis-à-vis the other survey types.}
    \label{fig:pop-unique}
\end{figure}

Figure \ref{fig:pop-unique} shows how $\Xi_\textbf{x}$ changes with the increasing number of attributes across 100 simulation runs. Naturally, the share constantly increases for the true survey. For the geomasked survey, the population uniqueness increases to a level of roughly 70\%. Recalling that the only difference between the geomasked survey and the true survey is the perturbed zip code, the remaining 30\% corresponds to the average number of survey respondents assigned to a new zip code due to the spatial anonymization process. Thus, not considering the zip code in the re-identification effort would let the population uniqueness of the geomasked survey also converge towards 1 similar to the true survey, even though requiring further knowledge on additional attributes. For the synthetic survey, the curve remains almost flat. The initial bump can largely be explained by the probability of a random combination of attributes representing an actual population unique in a small (area) sample size setting. Besides this theoretical argument, synthetic data always provides plausible deniability to the survey respondents. Similarly to our definition, \cite{Rocher2019EstimatingModels} use a Gaussian copula model to estimate the empirical likelihood of population uniqueness in incomplete datasets such as $D$ by assuming $\Xi_\textbf{x} \sim \text{Binomial}(\mathbb{1}_{\textbf{x}^{\{i\}}}, n_D)$ with $\textbf{x}^{\{i\}} i.i.d.$. While this approach is an excellent alternative to measure the re-identification risk in micro-level survey data when no validation data (in our experiment the 2011 Costa Rican census) is available, it assumes that the individual records are independent and identically distributed, which may be contestable in the presence of hierarchical dependencies and complex sampling designs.

Therefore, Figure \ref{fig:pop-unique} gives a strong indication that geomasking provides little additional safeguards for the respondents' privacy compared to the true survey in the presence of third-party information on a subset of the contained attributes. Hence, we consider an alternative microdata dissemination strategy: instead of publishing original microdata with perturbed cluster locations, we investigate the option of publishing two datasets - original microdata stripped of geographic identifiers below the level of strata and synthetic microdata with the original cluster locations. The choice is motivated by adopting a user-centric perspective: official household survey publications predominantly report on results up to the strata-level as results below are usually considered not representative. Analysis that benefits from below strata-level data often investigates proximity-related questions such as distances to certain locations and surrounding habitat. For the former, cluster locations are of minor importance, for the latter, however, the perturbation procedure introduces significant levels of uncertainty to the analysis \citep{warren2016influence}. The alternative microdata dissemination strategy obviously conserves data utility for analysis on the representative level via the first dataset, while the second dataset allows for the accurate capture of proximity-related information. However, two potential shortcomings need to be considered: first, can we use the synthetic dataset to predict the `private' attribute in the original dataset, i.e. the small area identifier, thus bypassing the privacy protection measures? Second, is the uncertainty we introduce by synthesizing the non-spatial attributes for spatial analysis smaller than the uncertainty from perturbing the cluster locations?

\subsection{Risk of re-identifying private geocodes}

\begin{figure}[ht!]
    \centering
    \includegraphics[width=\textwidth]{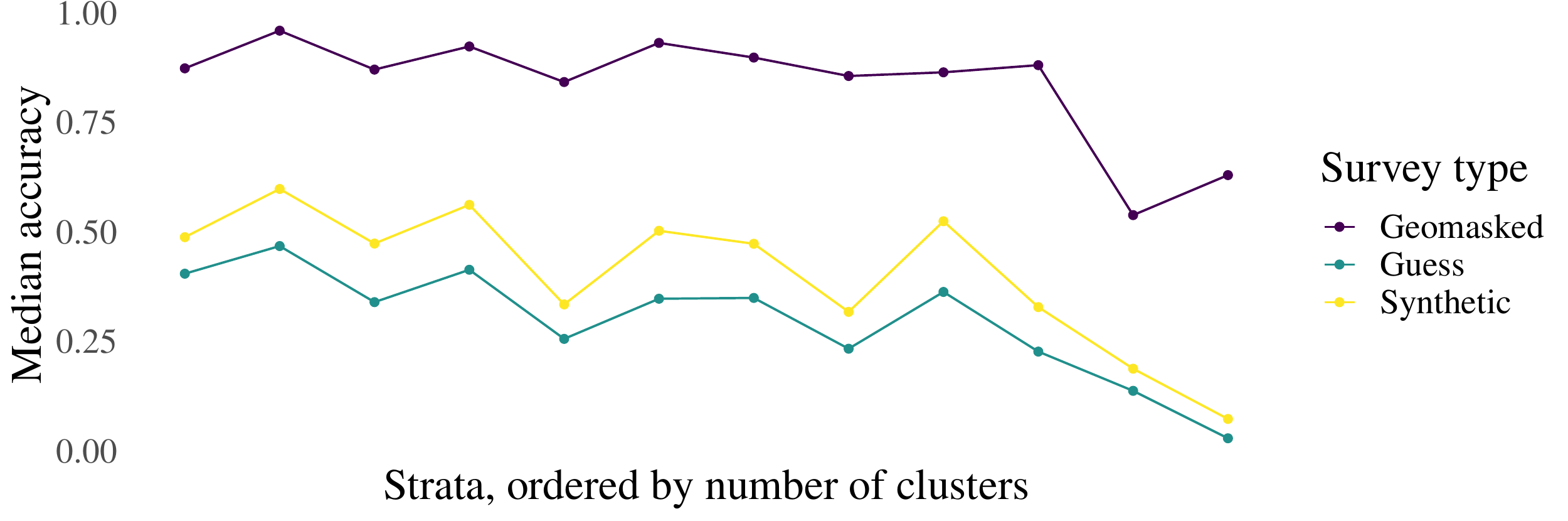}
    \caption{\textbf{Re-identification of the zip code as private attribute in the true survey for each stratum across 100 simulation runs.}\\
    Accuracy is measured by the share of successfully re-identified zip code labels in the true survey. A random forest model is trained on perturbed data, i.e. the geomasked and the synthetic survey, respectively. We evaluate the results against the true zip code labels in the true survey and compare them against random guesses of the private attribute.}
    \label{fig:acc-sens}
\end{figure}

To investigate the first shortcoming, we train a random forest model on the small area identifier - the zip code - in the anonymized surveys for each stratum separately. We use the trained models on the original data to predict the zip code for each record. Finally, we evaluate our predicted label against the original label. In addition, we compare the outcomes to randomly guessing the correct label in order to account for the number of small areas within each stratum. Figure \ref{fig:acc-sens} shows the median accuracy of the approaches across 100 simulation runs. While we are able to successfully re-construct the original zip code in most cases for the geomasked survey, it does not work much better for the synthetic data than for the random guess.

In our experiment, only one stratum consequently hosts more than ten small areas across all simulation runs, with one stratum hosting only two small areas in some simulation runs, giving the random guess also a good chance to predict correctly. Recalling from Figure \ref{fig:pop-unique} that roughly 70\% of the displaced clusters stay within the same zip code in the geomasked survey, even predicting the sensitive attribute for strata hosting as little as two small areas, average population uniqueness in the synthetic data would not exceed much the 50/50-chance of the random guess\%, thus providing better privacy protection in the re-identified original survey than the geomasked alternative.
%Thus - from a privacy perspective - the copula-based synthetic data approach outperforms the geomasked approach.

\subsection{Utility for survey augmentation}

\begin{figure}[ht!]
\captionsetup[subfigure]{justification=centering}
     \centering
     \begin{subfigure}[t]{0.24\linewidth}
         \centering
         \includegraphics[width=\textwidth]{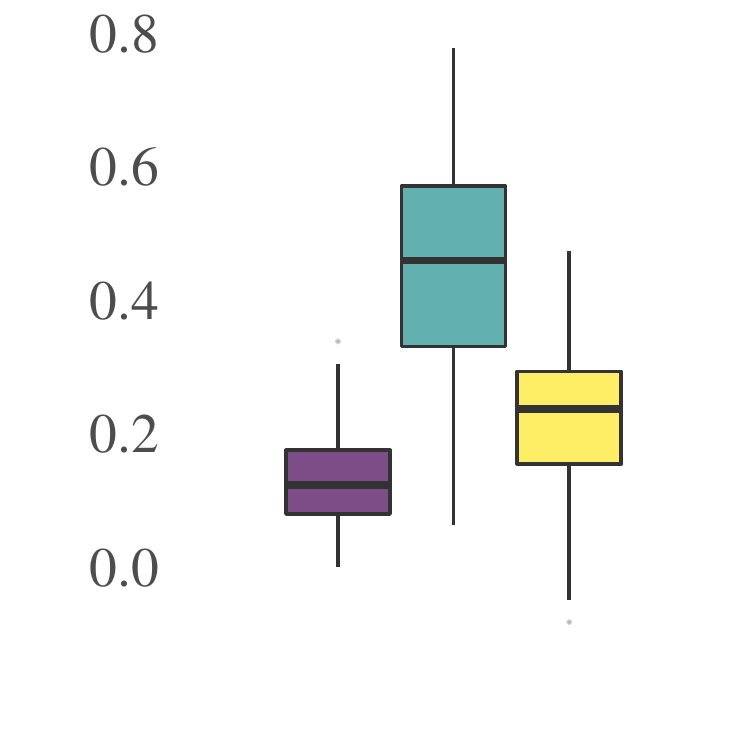}
         \caption{Adjusted R2}
         \label{fig:pred-adj}
     \end{subfigure}
     \begin{subfigure}[t]{0.24\linewidth}
         \centering
         \includegraphics[width=\textwidth]{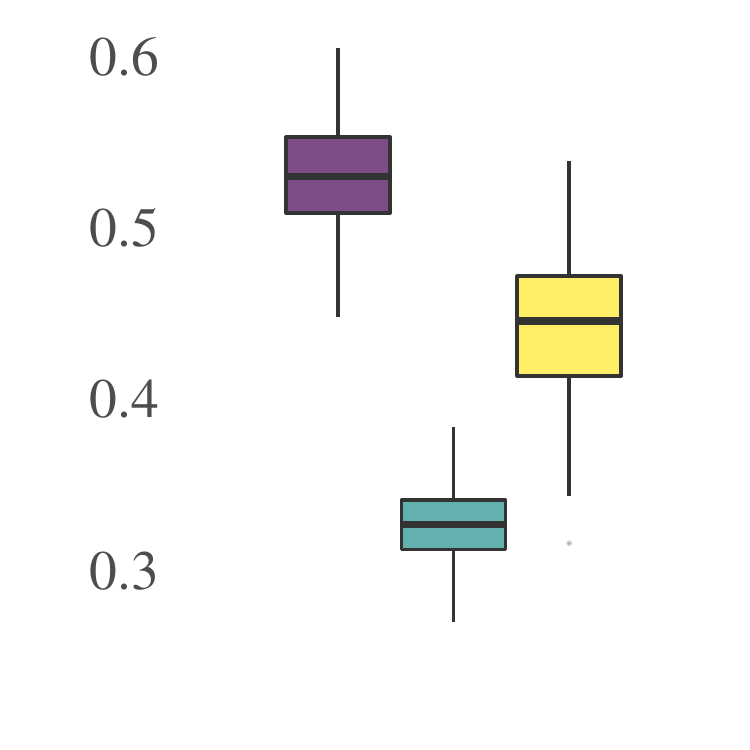}
         \caption{Relative Bias}
         \label{fig:pred-bias}
     \end{subfigure}
     \begin{subfigure}[t]{0.24\linewidth}
         \centering
         \includegraphics[width=\textwidth]{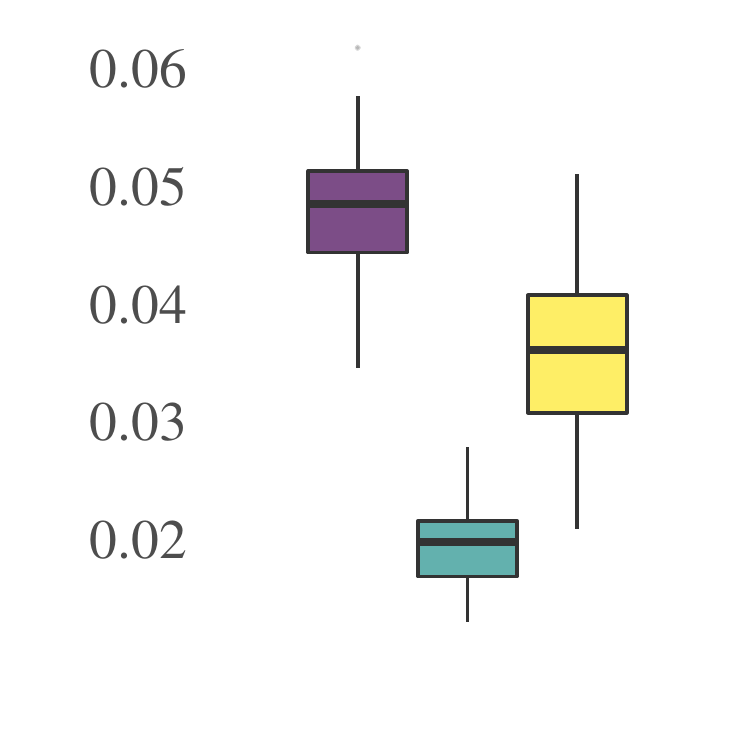}
         \caption{MSE}
         \label{fig:pred-mse}
     \end{subfigure}
     \begin{subfigure}[t]{0.15\linewidth}
         \centering
         \includegraphics[width=\textwidth]{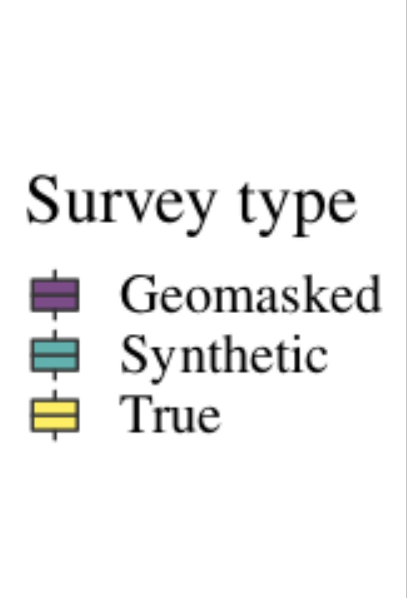}
        %  \caption{MSE}
        %  \label{fig:pred-densities}
     \end{subfigure}
     \hfill
     \begin{subfigure}[b]{\linewidth}
         \centering
         \includegraphics[width=\textwidth]{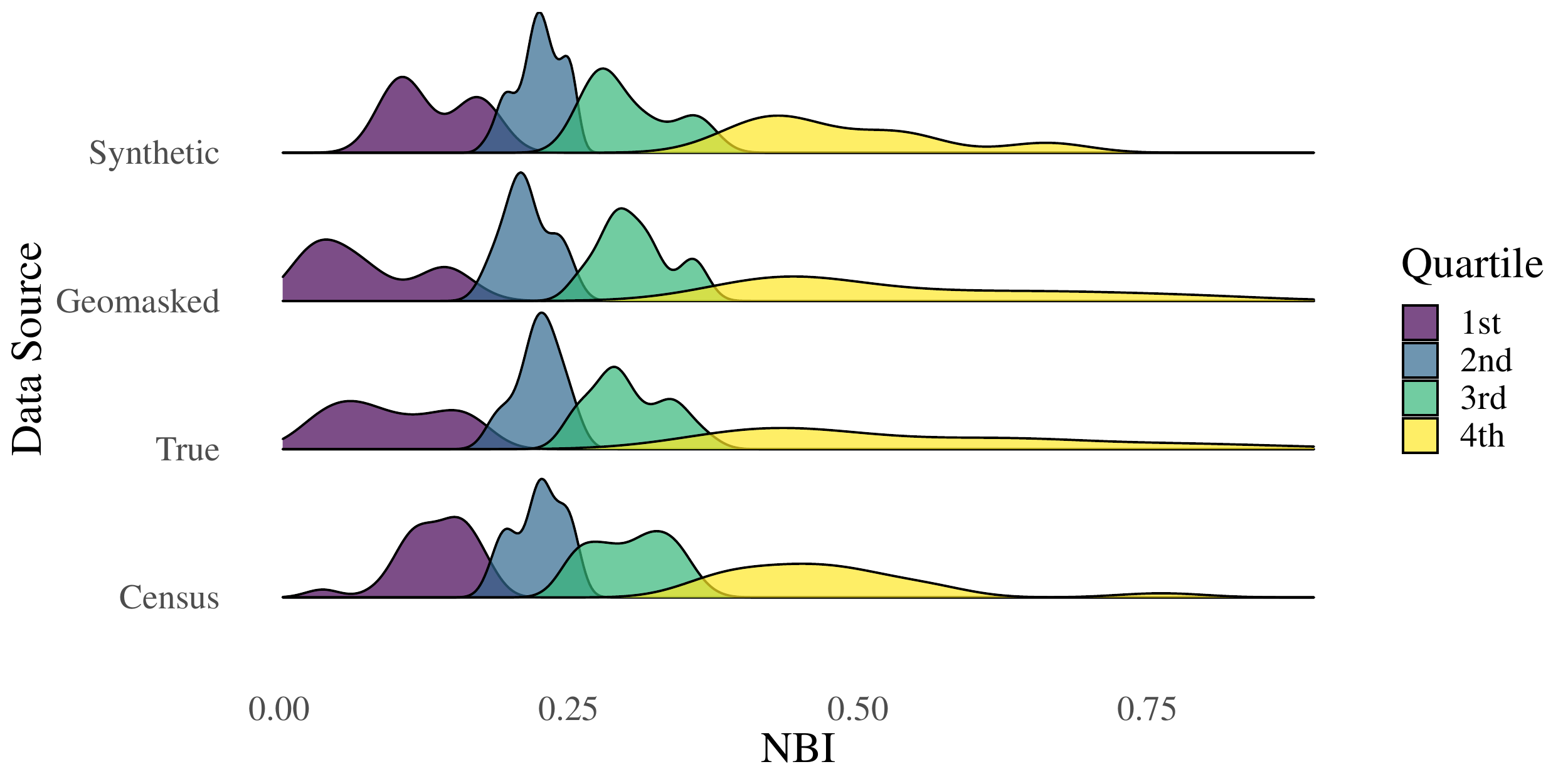}
         \caption{NBI densities}
         \label{fig:pred-densities}
     \end{subfigure}
        \caption{\textbf{Performance metrics of survey-based NBI estimates on the zip code-level.}\\ (a) Adjusted $R^2$ is based on the in-sample zip codes. (b) and (c) are based on the full sample and predictions are evaluated against the census across 100 simulation runs. (d) compares zip code-level NBI averages for a single simulation run.}
        \label{fig:pred}
\end{figure}

To give an indication about the utility of the different anonymization approaches, we use a setup common in recent academic literature (cf. \cite{Pokhriyal2017, leasure2020national, Schmid2017ConstructingSenegalb}): we augment the surveys with auxiliary information from geospatial (big) data. Specifically, we construct zip code-level aggregates from gridded satellite-derived features available from the WorldPop repository \citep{WorldPopwww.worldpop.org-SchoolofGeographyandEnvironmentalScienceUniversityofSouthamptonDepartmentofGeographyandGeosciencesUniversityofLouisvilleDepartementdeGeographie2018WorldPopb1} and combine them with zip code-level survey aggregates to provide predictions, especially for areas not sampled in the survey. As our target variable, we select the Unsatisfied Basic Needs index (\textit{Necesidades Básicas Insatisfechas (NBI)}) - a composite indicator similar to the multidimensional poverty index (MPI) \citep{Mendez2011, Alkire2019The2019} used as a key statistical indicator in Costa Rica. Details on the index can be found in the Supplementary Information. We evaluate our predictions against the census in terms of adjusted $R^2$, bias and the Mean Squared Error (MSE). Figures \ref{fig:pred-adj} - \ref{fig:pred-mse} show the performance along these three evaluation criteria across 100 simulation runs.

 Surprisingly, the synthetic approach not only outperforms the geomasked survey, it also provides predictions more in line with the census results than the true survey. A possible explanation could be that the copula approach reduces the impact of outliers on the zip code-specific NBI sample averages. This explanation is supported by Figure \ref{fig:pred-densities} that shows the distribution of zip code-level NBI averages grouped into quartiles for one simulation run as both the synthetic survey and the census showcase smaller tails in their distributions, respectively.
 %Table \ref{tab:pred-best-of} gives an overview about the best-performing approach by simulation round and survey type.
% \begin{table}[ht]
% \centering
% \small{
%     \begin{tabular}{lrrr}
%       \hline
%      Survey type & Adjusted $R^2$ & Relative Bias & RMSE \\ 
%       \hline
%     True & 2 & 10 & 9 \\ 
%     Geomasked & 0 & 0 & 0 \\ 
%     Synthetic & 98 & 90 & 91 \\ 
%       \hline
%     \end{tabular}
% 	\caption{\textbf{Best performing survey type by simulation round (100 rounds in total)}}
% 	\label{tab:pred-best-of}
% 	}
% \end{table}
We run additional experiments to compare the directly synthesized NBI and its underlying indicators with their counterparts computed from synthetic survey variables (see Supplementary Table 2 %\ref{table:appendix-pred} 
and Supplementary Fig. 4 %\ref{fig:appendix-nbi} 
in the Supplementary Information).

\section{Discussion}

In this paper, we proposed and evaluated an alternative data dissemination strategy for micro-level survey data that improves the trade-off between privacy risk and data utility. Specifically, we showed that by publishing two datasets, namely the original survey data with limited geographic identifiers and a synthetically-generated survey dataset with the true cluster locations, re-identification risks can be reduced significantly vis-à-vis popular geomasking approaches without incurring additional losses in terms of data utility for survey augmentation. This could help mapping initiatives such as \hyperlink{https://www.worldpop.org/}{WorldPop} or \hyperlink{https://grid3.org/}{GRID3} to improve their products as more accurate spatial data is available. In addition, by separating the marginals from the dependence structure, it provides data producers such as National Statistical Offices also with a useful tool to update the respective synthetic microdata files for the following years by updating the margins with nationally representative new data as sub-nationally representative surveys may only be conducted every few years. In the Supplementary Information, we further investigate the stability of our results by alternating the experiment design. 

First, while we chose the strata for the main analysis as they provide 'large-enough' sample sizes at the same time explicitly accounting for at least high-level regional variation, we study in further experiments whether fitting on smaller or larger geographic levels may better capture local variation at the expense of running into the risk of small sample problems or vice versa. Supplementary Fig. 1 %\ref{fig:appendix-sens-geo} 
summarizes the results for our copula model being fitted on the whole survey, the twelve strata and the zip code-level, respectively. It shows that by selecting the strata as our fitting level, we strike a balance between the underlying sample size (usually the larger the better) and capturing regional variation (usually the more disaggregated the better) both in terms of utility and risk. In addition, by using subsets of the full microdata for model fitting, the approach becomes computationally tractable also for larger surveys. 

Second, since generative models allow us to sample an arbitrary number of synthetic observations, we look at the impact of the synthetic sample size on the outcomes of the survey augmentation experiment, notably the adjusted $R^2$ and a measure of confidence in the direct survey estimates of the Fay-Herriot model (cf. Section \ref{section:methods-fh}) - the shrinkage factor $\gamma$. Supplementary Fig. 2 %\ref{fig:appendix-gamma} 
shows that with an increasing sample size, $\gamma$ increases as well, thus shifting more weight to the direct estimate. Even though intuitive as the sampling variance naturally decreases in $n_D$, at some point it may become misleading with potentially negative effects on the model performance as the synthetic data generating process still relies on the same information conveyed in the true survey with sample size $n_D$. However, in our experiment the adjusted $R^2$ does not exhibit a bump, but increases monotonically, thus hinting at little additional explanatory power of our satellite-derived covariates vis-à-vis the area-level direct survey estimate for the in-sample areas.

Third, since our target variable \textit{NBI} is a composite indicator, we compare the different composition levels of the synthetic NBI with the NBI constructed from synthetic data. While the divergence measure shows an overall good fit for the underlying indicators (see Supplementary Table 2 %\ref{table:appendix-pred} 
and Supplementary Fig. 4), %\ref{fig:appendix-nbi}
correlations are low, especially for higher-level compositions as the dimensions or the NBI itself.

Lastly, we test alternative encoding schemes for the transformation of categorical data. Also, we relax our assumption of the normally distributed margins by opening up to a wider group of parametric copulas (such as beta, gamma or uniform distributions) selected for each margin individually based on the two-sample Kolmogorov-Smirnov (KS) statistic to study the effect of the specification choice on the normalized KL divergence. Supplementary Fig. 3 %\ref{fig:appendix-encoding-marginals} 
shows that neither the encoding scheme nor the specification of the marginal distributions have large effects on the quality of the synthetically generated data.

Nevertheless, our approach is not without limitations. The copula-based approach towards synthetic data generation largely fails to correctly capture lower-level hierarchical relationships such as \textit{individuals - line numbers - households - houses} from the original data. As said before, since we see our analysis using a na\"ive Gaussian copula model as providing somewhat a lower bound for improving the utility-risk trade-off by adopting the proposed microdata dissemination strategy vis-à-vis common geomasking approaches, there is much room for improvement. To name a few, latent copula designs can be considered to avoid data transformations, marginal distributions can be modelled non-parametrically, hierarchical structures can be accounted for more rigorously by either modelling the hierarchies separately as suggested by \cite{templ2017SDC} or by modelling the relationships explicitly.
%In addition, deep learning approaches such as CTGAN \citep{Xu2019} can be applied to better capture non-linear relationships from the original data. 
In addition, synthetic data may - under some circumstances - leak private information, e.g. through the generated value ranges. As a response, differentially-private implementations of existing generative models have been proposed such as PrivBayes \citep{zhang2017}, PrivSyn \citep{Zhang2021} and PATE-GAN \citep{Jordon2019}.

%[What about hierarchical dependencies? What about large p settings? Is there a systematic bias how minority classes are captured? What about use cases where you use the actual distance?]

%Retaining full data utility while offering 
%[Here, we need to stress the impact of our proposal beyond our Costa Rican experiment, we need to flag the shortcomings of the approach (e.g. capturing lower-level hierarchies in the data, never really being able to fully assess utility), we need to give a short overview/summary about what we do in the Supplementary Information (different copula fitting options, different sample sizes, additional goodness-of-fit tests) and finally lay out a path for future research.]

\section{Methods}\label{section:methods}

\subsection{Fitting Gaussian copulas to survey attributes}

As an alternative to geomasking, we use synthetically generated survey attributes for protecting the respondents' privacy while keeping the true point locations of the selected clusters. To do so, we fit a Gaussian copula model on the original survey attributes $\textbf{x}$ and sample from the learned joint distribution for each cluster individually with the originally sample size $n_j$. Therefore, consider our survey $D$, where $X_1$ represents a random variable with a continuous marginal cumulative distribution function (cdf) denoted by $F_{1}(x_1) = P(X_1\leq x_1)$.
%Consider also $F_{x_{1} x_{2}}(x_{1} x_{2}) = P(X_{1}\leq x_1, X_{2}\leq x_2)$ being the joint cumulative distribution function of $(X_1,X_2)$.
For the multivariate case, the joint cdf can be generalized to $F_{1, \dots ,d}(x_{1}, \dots , x_{d}) = P(X_{1}\leq x_1, \dots,  X_{d}\leq x_d)$.

A copula, firstly introduced in the work of \cite{sklar1959}, is a cumulative density function with uniform marginals between [0,1]. Thus - based on Sklar´s theorem \citep{sklar1959} - when all variables are continuous, the d-dimensional random vector $X_1, \dots, X_d$ can be defined in a uniform space $[0,1]^d$, creating a random vector $U_1, \dots, U_d$ via the probability integral transform $u_d = F_d(x_d)$. In this case, a unique d-dimensional copula $C(u_d)$ exists:

\begin{equation}
    C(u_1, \dots, u_d) = F\big(F_{1}^{-1}(u_1), \dots , F_{d}^{-1}(u_d)\big)
\end{equation}

As motivated in Section \ref{section:results-copula}, we account for the fact that household surveys largely consist of categorical variables by applying data transformation. Among the plethora of possible encoding schemes, the most common encoding scheme is one-hot encoding, where for each class of a categorical variable a binary dummy variable is created \citep{Benali2021}.
% Algorithms \ref{alg:three} and \ref{alg:four} provide details.
%
% \normalem %%%% disable auto underline
%
% \begin{minipage}[t]{0.45\textwidth}
%   \vspace{0pt}  
%     \begin{algorithm}[H]
%         \caption{Transform nominal variables}\label{alg:three}
%         \textbf{Input $D = (X_1, \dots, X_d)$} \\
%         \textbf{Output $\tilde{D} = (\tilde{X}_1, \dots, \tilde{X}_m)$}
%         \BlankLine
%       
%         \For{$X_d \in D$}{
%                   \If{$X_d$ is Continuous}{
%           $\tilde{X}_p \gets X_d$}
%           \If{$X_d$ is Non-continuous}{
%           $\tilde{\textbf{X}}_q =  (X_{d_1}, \dots, X_{d_v}) \gets \mathrm{T}(X_{d})$}
%         }
%         $\tilde{D} \gets (\tilde{X}_p, \tilde{\textbf{X}}_q)  \quad \forall \quad p \in P \text{ and }q \in Q$
%     \end{algorithm}
% \end{minipage}
% \begin{minipage}[t]{0.45\textwidth}
%  
%   \vspace{0pt}  
%  
% \begin{algorithm}[H]
% \caption{Back-transform dummies}\label{alg:four}
% \textbf{Input $\tilde{S} =  (\tilde{Y}_1, \dots, \tilde{Y}_m)$} \\
% \textbf{Output $S = (Y_1, \dots, Y_d)$}
% \BlankLine
%
% \For{$\tilde{Y}_m \in \tilde{S}$}{
%           \If{$\tilde{Y}_m$ is not indexed as class of $q$}{
%   $Y_p \gets \tilde{Y}_m$}
%   \If{$\tilde{Y}_m$ is indexed as class of $q$}{
%   $Y_{q_v} \gets \tilde{Y}_{m}$}
%
% %  $Y_d \gets (Y_p \oplus Y_q)$ 
% }
%
% \For{$q \in Q$}{
%  $Y_q \gets \mathrm{T}^{-1}(Y_{q_1}, \dots, Y_{q_v})$
% }
%
% %  $S \gets (\textbf{Y}_p \oplus \textbf{Y}_q)$\\
%  $S \gets (Y_p, Y_q) \quad \forall \quad p \in P \text{ and }q \in Q$ 
%
%
% \end{algorithm}
% \end{minipage}
%
% \ULforem %%%% enable auto underline
%
A disadvantage of this option is that it may become computationally challenging and prone to multicollinearity in the presence of variables with a high cardinality, i.e. with a large number of classes, since each possible class creates a new variable \citep{Bourou2021}. %Even though other approaches such as assigning ordinal levels to nominal classes exist, we opt for one-hot encoding because it incurs no information loss and requires no additional testing [CITATION]. The algorithms to transform categorical attributes to dummy variables and back prior and after applying the copula model are described in Algorithm \ref{alg:three} and \ref{alg:four}, respectively. 
Interestingly, there is -- to the best of our knowledge -- little comprehensive, comparative and conclusive scientific evidence on the properties and performance of different categorical encoding schemes. Therefore, we explore two other well-known alternatives with more favourable computation times: ordinal and frequency encoding. Ordinal encoding uses integers to represent each classes, e.g. from 0 to $v$, the number of classes in a categorical variable. Assigning an unreal order to nominal variables is the main pitfall of this alternative \citep{jiangdan2020}. Frequency encoding -- as used in medical imaging \citep{mansfield1977medical} and similar to the concept of \textit{term frequency} in Natural Language Processing \citep{aizawa2003information} -- assigns an interval in [0,1] to each class based on and ordered by its proportion of occurrence. Then, it uses the middle point of each interval as float representative of the respective class. Back-transformation is done by assigning a new point to a class via the respective interval it falls into. In this sense, this alternative conveys information of the importance of each class \citep{Sabharwal2021}. Based on the results of the different encoding schemes shown in Supplementary Fig. 3, %\ref{fig:appendix-encoding-marginals} 
we opt for the frequency encoding scheme in the following. Algorithms \ref{alg:three} and \ref{alg:four} provide details on the chosen scheme.

\normalem %%%% disable auto underline

\begin{minipage}[t]{0.45\textwidth}
  \vspace{0pt}  
    \begin{algorithm}[H]
        \caption{Transform categorical variables}\label{alg:three}
        \textbf{Input $D = (X_1, \dots, X_d)$} \\
        \textbf{Output $\tilde{D} = (\tilde{X}_1, \dots, \tilde{X}_d)$}
        \BlankLine
      
        \For{$X_d \in D$}{
                  \If{$X_d$ is Continuous}{
          $\tilde{X}_p \gets X_d$}
          \If{$X_d$ is Non-continuous}{
        %   $\tilde{\textbf{X}}_q =  (X_{d_1}, \dots, X_{d_v}) \gets \mathrm{T}(X_{d})$}
          $\tilde{X}_q \gets  \mathrm{T}(X_d)$}
        }
        $\tilde{D} \gets (\tilde{X}_p, \tilde{X}_q)  \quad \forall \quad p \in P \text{ and }q \in Q$
    \end{algorithm}
\end{minipage}
\begin{minipage}[t]{0.45\textwidth}
 
  \vspace{0pt}  
 
\begin{algorithm}[H]
\caption{Back-transform frequency encoded variables}\label{alg:four}
\textbf{Input $\tilde{S} =  (\tilde{Y}_1, \dots, \tilde{Y}_d)$} \\
\textbf{Output $S = (Y_1, \dots, Y_d)$}
\BlankLine

\For{$\tilde{Y}_d \in \tilde{S}$}{
          \If{$\tilde{Y}_d$ is not indexed as variable in $Q$}{
  $Y_p \gets \tilde{Y}_d$}
  \If{$\tilde{Y}_d$ is indexed as variable in $Q$}{
  $Y_{q} \gets T^{-1}(\tilde{Y}_{d})$}

%  $Y_d \gets (Y_p \oplus Y_q)$ 
}

% \For{$q \in Q$}{
%  $Y_q \gets \mathrm{T}^{-1}(Y_{q_1}, \dots, Y_{q_v})$
% }

%  $S \gets (\textbf{Y}_p \oplus \textbf{Y}_q)$\\
 $S \gets (Y_p, Y_q) \quad \forall \quad p \in P \text{ and }q \in Q$

\end{algorithm}
\end{minipage}

\ULforem %%%% enable auto underline

Thus, the d-dimensional Gaussian copula $C_{\Sigma}^{G}(\textbf{u})$ is defined as the cdf of a multivariate normal distribution $\mathcal{N}(\mu, \Sigma)$ with $\Sigma \in \mathbb{R}^{dxd}$ represented on the unit cube $[0,1]^d$: 
\begin{equation}
    C_{\Sigma}^{G}(u_1, \dots, u_d) = \phi_{\Sigma}\big(\phi^{-1}(u_1), \dots , \phi^{-1}(u_d)\big)
\end{equation}

The density of a Gaussian copula is then defined as: 

\begin{equation}
c_{\Sigma}^{G}(\textbf{u}) = \frac{1}{\sqrt{\text{det }\Sigma}} \text{exp}\Big( -\frac{1}{2} \phi ^{-1} (\textbf{u}) ^T \cdot (\Sigma ^{-1}-I) \cdot\phi ^{-1} (\textbf{u}) \Big)
\end{equation}

with $\textbf{u} \in [0,1]^d$, $I \in \mathbb{R}^{dxd}$ being the identity matrix, and $\phi ^{-1}$ being the inverse cumulative distribution function of a standard normal distribution. $\Sigma$ is a positive semi-definite covariance matrix that we estimate based on Pearson's correlation coefficient $\rho$ \citep{Li2014}.

As noted in Section \ref{section:results-copula}, we sample for each cluster individually with a sample size of $n_j$. While rejection sampling could be an option for ensuring only synthetic rows with the respective cluster identifier are selected, it proves computationally inefficient. With copulas being multivariate cdfs, we introduce conditions instead. Hence, we sample from a multivariate normal distribution conditional on cluster $j$. Thus, our transformed dataset $\Tilde{D}$ with one conditional variable becomes $\Tilde{D} = (\Tilde{\textbf{X}}_a|\Tilde{\textbf{X}}_{b})$ with $ \Tilde{\textbf{X}}_a := \Tilde{X}_1, \dots ,\Tilde{X}_{d-1}$ being the transformed attributes to be synthesized and $\Tilde{\textbf{X}}_{b} := \Tilde{X}_{d}$ being the transformed cluster identifier. The parameters of the respective multivariate normal distributions are thus partitioned into:

\begin{equation}
\Tilde{D}
=
\begin{bmatrix}
\Tilde{\textbf{X}}_a\\
\Tilde{\textbf{X}}_b
\end{bmatrix}
\text{,} \textcolor{white}{x}
\mu =
\begin{bmatrix}
\mu_{a}\\
\mu_{b}
\end{bmatrix}
\text{and} \textcolor{white}{x}
\Sigma =
\begin{bmatrix}
\Sigma_{a a}&\Sigma_{a b}\\
\Sigma_{b a} &\Sigma_{b b}
\end{bmatrix} 
\end{equation}

with $\mu_{a} \in \mathbb{R}^{d-1}$ and $\mu_{b} \in \mathbb{R}^1$ and $\Sigma_{a a} \in \mathbb{R}^{(d-1) \times (d-1)}$, $\Sigma_{a b} \in \mathbb{R}^{(d-1) \times 1}$,  $\Sigma_{b a} \in \mathbb{R}^{1 \times (d-1)}$, and $\Sigma_{b b} \in \mathbb{R}^{1 \times 1}$ being the means and positive semi-definite covariance matrices, respectively. Following Algorithm \ref{alg:two}, the parameters of our estimated marginal distributions $\Psi$ and of the copula $C_{\Sigma}^{G}(\textbf{u})$ need to be adapted to mirror the conditionality such that $\Psi_{a|b}(\Tilde{\textbf{X}}_a|\Tilde{\textbf{X}}_b)$ and $C_{\Sigma}^{G}(\textbf{u}_a|\textbf{u}_{b})$.
%such that $(\Tilde{\textbf{X}}_a, \Tilde{\textbf{X}}_b) \sim \mathcal{N}((x_a, x_b); \mu_{a, b}, \Sigma_{a, b})$
%such that $\psi(X_{a|b = k} \sim \mathcal{N}(\mu_a + \frac{\sigma_1}{\sigma_1}*\rho_{a,b}(k-\mu_b), (1-\rho_{a,b})*\sigma^2_a)$ with $\mu = $. the parameters of a conditional multivariate normal distribution must be specified. Let $X = (X_1, \dots ,X_d)$ \textcolor{red}{or $\Tilde{D} = (\Tilde{X}_1, \dots ,\Tilde{X}_m )$} be a $m-$dimensional random vector with the following distribution $X \sim \mathcal{N}(\mu, \Sigma)$, with $\mu$ the mean vector and $\Sigma$ the $m \times m$ covariance matrix. The vector $X$ can be partitioned in $X_1$ and $X_2$ %to indicate that the distribution of $X_1$ is conditional on $X_2$, now $l+m-$ dimensional. Then,
% then, $(X_1, X_2) \sim \mathcal{N}((x_1, x_2); \mu_{x_1, x_2}, \Sigma_{x_1, x_2})$, with
Consequently, we sample from $\sim \mathcal{N}(\Bar{\mu}, \Bar{\Sigma}$)
%$x_1$ conditional on $x_2$ is now defined as $(X_1|X_2) \sim \mathcal{N}(x_1; \mu_{x_1|x_2}, \Sigma_{x_1|x_2})$
with: 

\begin{equation}
    \Bar{\mu} = \mu_{a} + \Sigma_{a b}\Sigma^{-1}_{b b} (X_b -\mu_b) \in \mathbb{R}^{d-1}
\end{equation}

and

\begin{equation}
    \Bar{\Sigma} =  \Sigma_{a a} -  \Sigma_{a b} \Sigma^{-1}_{b b} \Sigma_{b a} \in \mathbb{R}^{(d-1) \times (d-1)}.
\end{equation}

We iterate the copula-based fitting and sampling procedure for every stratum separately as it allows to better capture sub-national variation using representative sub-samples and as it proves computationally more tractable. For sampling designs with varying household- or individual-level inclusion probabilities (e.g. in the DHS, women - in comparison to men - are usually oversampled), \cite{templ2017SDC} suggests to sample a synthetic population and re-iterate the sampling procedure to produce valid synthetic sampling weights. As in our design sampling weights are identical across households for a given PSU due to the systematic sampling approach in the second stage, the original sampling weights remain valid. The virtue in our model choice is the relative simplicity, little requirements in terms of ex-ante knowledge about the individual distributions $X_d$ and its computational efficiency. For further experiments on the robustness and sensitivity of our modelling choices, we refer to the Supplementary Information.

%\begin{comment}

%\end{comment}

\subsection{Area-level survey augmentation methods}\label{section:methods-fh}

Survey data can be augmented with the use of area-level models, e.g. the Fay-Herriot model \citep{fay1979estimates} by linking direct estimators gathered from survey data to relevant auxiliary information. Both, direct estimators, and auxiliary data are aggregated on $k=1, \dots, D$ areas. Traditionally, these auxiliary covariates $\textbf{x}_k$ are obtained from recent censuses, administrative records or other geospatial (big) data sources. In this paper, we make use of satellite imagery features as area-level covariates. The Fay-Herriot is a two-level model, the first part is composed by the sampling model:

\begin{equation}
    \hat{\theta}^{\text{Dir}}_k =  \theta_k + e_k, \hspace{5mm} e_k \sim N(0,\sigma^2_{e_k}),
\end{equation}

where the sampling error is represented by $e_k$ and $\hat{\theta}^{\text{Dir}}_k$ is the direct estimator of $\theta_k$ (e.g. sample mean). The linking model provides the second part, where relevant area-level covariates are considered:

\begin{equation}
    \theta_k = \textbf{x}'_k \hat{\beta} + u_k.
\end{equation}

Here, the random area effects $u_k$ are assumed to be independent with mean 0 and variance $\sigma^2_u$. The empirical best linear unbiased predictor (EBLUP) estimator is given by:

\begin{equation}
    \hat{\theta}^{\text{FH}}_k = \gamma_k \hat{\theta}^{Dir}_k + (1- \gamma_k) \textbf{x}'_k \hat{\beta} =  \textbf{x}'_k \hat{\beta} + \hat{u}_k, \textcolor{white}{............}
\end{equation}

with $\gamma_k = \frac{\hat{\sigma}^2_u}{\hat{\sigma}^2_u+\hat{\sigma}^2_{e_k}}$ denoting the shrinkage factor for each area $k$. The parameter estimates of this model can be obtained via maximum likelihood (ML) or restricted ML (REML). Note that the shrinkage factor allows to weight in favor of the direct estimator when sampling variances are small; on the contrary the synthetic estimator $\textbf{x}'_k \hat{\beta}$ receives more weight when the sampling variance is larger. Results on an experiment studying the sensitivity of the shrinkage factor and adjusted $R^2$ for varying synthetic sample sizes are shown in Supplementary Fig. 2. %\ref{fig:appendix-gamma} 
Further details on the Fay-Herriot model can be found in \cite{RaoMol15}.

\newpage

\bibliography{refs}

\begin{thebibliography}{10}
\expandafter\ifx\csname url\endcsname\relax
  \def\url#1{\texttt{#1}}\fi
\expandafter\ifx\csname urlprefix\endcsname\relax\def\urlprefix{URL }\fi
\providecommand{\bibinfo}[2]{#2}
\providecommand{\eprint}[2][]{\url{#2}}

\bibitem{granello2004}
\bibinfo{author}{Granello, D.~H.} \& \bibinfo{author}{Wheaton, J.~E.}
\newblock \bibinfo{title}{Online data collection: Strategies for research}.
\newblock \textit{\bibinfo{journal}{J Couns Dev.}}
  \textbf{\bibinfo{volume}{82}}, \bibinfo{pages}{387--393}
  (\bibinfo{year}{2004}).
\newblock
  \urlprefix\url{https://onlinelibrary.wiley.com/doi/10.1002/j.1556-6678.2004.tb00325.x}.

\bibitem{Stevens2015DisaggregatingData}
\bibinfo{author}{Stevens, F.~R.}, \bibinfo{author}{Gaughan, A.~E.},
  \bibinfo{author}{Linard, C.} \& \bibinfo{author}{Tatem, A.~J.}
\newblock \bibinfo{title}{{Disaggregating census data for population mapping
  using Random forests with remotely-sensed and ancillary data}}.
\newblock \textit{\bibinfo{journal}{PLoS ONE}} \textbf{\bibinfo{volume}{10}},
  \bibinfo{pages}{e0107042} (\bibinfo{year}{2015}).
\newblock \urlprefix\url{https://doi.org/10.1371/journal.pone.0107042}.

\bibitem{leasure2020national}
\bibinfo{author}{Leasure, D.~R.}, \bibinfo{author}{Jochem, W.~C.},
  \bibinfo{author}{Weber, E.~M.}, \bibinfo{author}{Seaman, V.} \&
  \bibinfo{author}{Tatem, A.~J.}
\newblock \bibinfo{title}{{National population mapping from sparse survey data:
  A hierarchical Bayesian modeling framework to account for uncertainty}}.
\newblock \textit{\bibinfo{journal}{Proc. Natl. Acad. Sci. U.S.A.}}
  \textbf{\bibinfo{volume}{117}}, \bibinfo{pages}{24173--24179}
  (\bibinfo{year}{2020}).
\newblock \urlprefix\url{https://www.pnas.org/doi/abs/10.1073/pnas.1913050117}.

\bibitem{Pokhriyal2017}
\bibinfo{author}{Pokhriyal, N.} \& \bibinfo{author}{Jacques, D.~C.}
\newblock \bibinfo{title}{{Combining disparate data sources for improved
  poverty prediction and mapping}}.
\newblock \textit{\bibinfo{journal}{Proc. Natl. Acad. Sci. U.S.A.}}
  \textbf{\bibinfo{volume}{114}}, \bibinfo{pages}{E9783--E9792}
  (\bibinfo{year}{2017}).
\newblock
  \urlprefix\url{https://www.pnas.org/doi/full/10.1073/pnas.1700319114}.

\bibitem{Schmid2017ConstructingSenegalb}
\bibinfo{author}{Schmid, T.}, \bibinfo{author}{Bruckschen, F.},
  \bibinfo{author}{Salvati, N.} \& \bibinfo{author}{Zbiranski, T.}
\newblock \bibinfo{title}{{Constructing sociodemographic indicators for
  national statistical institutes by using mobile phone data: estimating
  literacy rates in Senegal}}.
\newblock \textit{\bibinfo{journal}{J. R. Stat. Soc. Ser. A Stat. Soc.}}
  \textbf{\bibinfo{volume}{180}}, \bibinfo{pages}{1163--1190}
  (\bibinfo{year}{2017}).
\newblock
  \urlprefix\url{https://rss.onlinelibrary.wiley.com/doi/abs/10.1111/rssa.12305}.

\bibitem{subash2018satellite}
\bibinfo{author}{Subash, S.~P.}, \bibinfo{author}{Kumar, R.~R.} \&
  \bibinfo{author}{Aditya, K.~S.}
\newblock \bibinfo{title}{{Satellite data and machine learning tools for
  predicting poverty in rural India}}.
\newblock \textit{\bibinfo{journal}{Agric. Econ. Res. Rev.}}
  \textbf{\bibinfo{volume}{31}}, \bibinfo{pages}{231--240}
  (\bibinfo{year}{2018}).
\newblock \urlprefix\url{https://ageconsearch.umn.edu/record/284254}.

\bibitem{fatehkia2020relative}
\bibinfo{author}{Fatehkia, M.}, \bibinfo{author}{Coles, B.},
  \bibinfo{author}{Ofli, F.} \& \bibinfo{author}{Weber, I.}
\newblock \bibinfo{title}{The relative value of facebook advertising data for
  poverty mapping}.
\newblock \textit{\bibinfo{journal}{Proc. Int. AAAI Conf. Web Soc. Media}}
  \textbf{\bibinfo{volume}{14}}, \bibinfo{pages}{934--938}
  (\bibinfo{year}{2020}).
\newblock
  \urlprefix\url{https://ojs.aaai.org/index.php/ICWSM/article/view/7361}.

\bibitem{Chi2022}
\bibinfo{author}{Chi, G.}, \bibinfo{author}{Fang, H.},
  \bibinfo{author}{Chatterjee, S.} \& \bibinfo{author}{Blumenstock, J.~E.}
\newblock \bibinfo{title}{{Microestimates of wealth for all low- and
  middle-income countries}}.
\newblock \textit{\bibinfo{journal}{Proc. Natl. Acad. Sci. U.S.A.}}
  \textbf{\bibinfo{volume}{119}}, \bibinfo{pages}{e2113658119}
  (\bibinfo{year}{2022}).
\newblock \urlprefix\url{https://www.pnas.org/doi/abs/10.1073/pnas.2113658119}.
\newblock \eprint{2104.07761}.

\bibitem{Blumenstock2018}
\bibinfo{author}{Blumenstock, J.~E.}
\newblock \bibinfo{title}{Estimating economic characteristics with phone data}.
\newblock \textit{\bibinfo{journal}{AEA Pap. Proc.}}
  \textbf{\bibinfo{volume}{108}}, \bibinfo{pages}{72--76}
  (\bibinfo{year}{2018}).
\newblock
  \urlprefix\url{https://www.aeaweb.org/articles?id=10.1257/pandp.20181033}.

\bibitem{Aiken2022}
\bibinfo{author}{Aiken, E.}, \bibinfo{author}{Bellue, S.},
  \bibinfo{author}{Karlan, D.}, \bibinfo{author}{Udry, C.} \&
  \bibinfo{author}{Blumenstock, J.~E.}
\newblock \bibinfo{title}{{Machine learning and phone data can improve
  targeting of humanitarian aid}}.
\newblock \textit{\bibinfo{journal}{Nature 2022}}
  \textbf{\bibinfo{volume}{603}}, \bibinfo{pages}{864–870}
  (\bibinfo{year}{2022}).
\newblock \urlprefix\url{https://www.nature.com/articles/s41586-022-04484-9}.

\bibitem{Grace2019}
\bibinfo{author}{Grace, K.} \textit{et~al.}
\newblock \bibinfo{title}{{Integrating Environmental Context into DHS Analysis
  While Protecting Participant Confidentiality: A New Remote Sensing Method}}.
\newblock \textit{\bibinfo{journal}{Popul. Dev. Rev.}}
  \textbf{\bibinfo{volume}{45}}, \bibinfo{pages}{197--218}
  (\bibinfo{year}{2019}).
\newblock
  \urlprefix\url{https://www.ncbi.nlm.nih.gov/pmc/articles/PMC6446718/}.

\bibitem{Brown2014}
\bibinfo{author}{Brown, M.~E.}, \bibinfo{author}{Grace, K.},
  \bibinfo{author}{Shively, G.}, \bibinfo{author}{Johnson, K.~B.} \&
  \bibinfo{author}{Carroll, M.}
\newblock \bibinfo{title}{{Using satellite remote sensing and household survey
  data to assess human health and nutrition response to environmental change}}.
\newblock \textit{\bibinfo{journal}{Popul. Environ.}}
  \textbf{\bibinfo{volume}{36}}, \bibinfo{pages}{48--72}
  (\bibinfo{year}{2014}).
\newblock \urlprefix\url{https://doi.org/10.1007/s11111-013-0201-0}.

\bibitem{arambepola2020spatiotemporal}
\bibinfo{author}{Arambepola, R.} \textit{et~al.}
\newblock \bibinfo{title}{Spatiotemporal mapping of malaria prevalence in
  madagascar using routine surveillance and health survey data}.
\newblock \textit{\bibinfo{journal}{Sci. Rep.}} \textbf{\bibinfo{volume}{10}},
  \bibinfo{pages}{18129} (\bibinfo{year}{2020}).
\newblock \urlprefix\url{https://doi.org/10.1038/s41598-020-75189-0}.

\bibitem{Koebe2020BetterModelling}
\bibinfo{author}{Koebe, T.}
\newblock \bibinfo{title}{{Better coverage, better outcomes? Mapping mobile
  network data to official statistics using satellite imagery and radio
  propagation modelling}}.
\newblock \textit{\bibinfo{journal}{PLoS ONE}} \textbf{\bibinfo{volume}{15}},
  \bibinfo{pages}{e0241981} (\bibinfo{year}{2020}).
\newblock \urlprefix\url{https://dx.plos.org/10.1371/journal.pone.0241981}.

\bibitem{armstrong1999geographically}
\bibinfo{author}{Armstrong, M.~P.}, \bibinfo{author}{Rushton, G.} \&
  \bibinfo{author}{Zimmerman, D.~L.}
\newblock \bibinfo{title}{Geographically masking health data to preserve
  confidentiality}.
\newblock \textit{\bibinfo{journal}{Stat. Med.}} \textbf{\bibinfo{volume}{18}},
  \bibinfo{pages}{497--525} (\bibinfo{year}{1999}).
\newblock
  \urlprefix\url{{https://onlinelibrary.wiley.com/doi/10.1002/\%28SICI\%291097-0258\%2819990315\%2918\%3A5\%3C497\%3A\%3AAID-SIM45\%3E3.0.CO\%3B2-\%23}}.

\bibitem{Kroll2016}
\bibinfo{author}{Kroll, M.} \& \bibinfo{author}{Schnell, R.}
\newblock \bibinfo{title}{{Anonymisation of geographical distance matrices via
  Lipschitz embedding}}.
\newblock \textit{\bibinfo{journal}{Int. J. Health Geogr.}}
  \textbf{\bibinfo{volume}{15}}, \bibinfo{pages}{1--14} (\bibinfo{year}{2016}).
\newblock \urlprefix\url{https://doi.org/10.1186/s12942-015-0031-7}.

\bibitem{West2017}
\bibinfo{author}{West, B.~T.} \textit{et~al.}
\newblock \textit{\bibinfo{title}{Establishing Infrastructure for the Use of
  Big Data to Understand Total Survey Error}}, chap.~\bibinfo{chapter}{21},
  \bibinfo{pages}{457--485} (\bibinfo{publisher}{John Wiley \& Sons, Ltd},
  \bibinfo{year}{2017}).
\newblock
  \urlprefix\url{https://onlinelibrary.wiley.com/doi/abs/10.1002/9781119041702.ch21}.

\bibitem{Dwork2008}
\bibinfo{author}{Dwork, C.}
\newblock \bibinfo{title}{Differential privacy: A survey of results}.
\newblock In \textit{\bibinfo{booktitle}{Theory and Applications of Models of
  Computation. TAMC 2008. Lecture Notes in Computer Science}}, vol.
  \bibinfo{volume}{4978}, \bibinfo{pages}{1--19}
  (\bibinfo{organization}{Springer, Berlin, Heidelberg}, \bibinfo{year}{2008}).
\newblock \urlprefix\url{https://doi.org/10.1007/978-3-540-79228-4_1}.

\bibitem{AndresGeo-Indistinguishability:Systems}
\bibinfo{author}{Andr{\'{e}}s, M.~E.}, \bibinfo{author}{Bordenabe, N.~E.},
  \bibinfo{author}{Chatzikokolakis, K.} \& \bibinfo{author}{Palamidessi, C.}
\newblock \bibinfo{title}{Geo-indistinguishability: Differential privacy for
  location-based systems}.
\newblock In \textit{\bibinfo{booktitle}{Proc. 2013 ACM SIGSAC Conf. Comput.
  Commun. Secur.}}, \bibinfo{pages}{901--914} (\bibinfo{year}{2013}).
\newblock \urlprefix\url{https://doi.org/10.1145/2508859.2516735}.

\bibitem{templ2017SDC}
\bibinfo{author}{Templ, M.}
\newblock \textit{\bibinfo{title}{Statistical disclosure control for
  microdata}} (\bibinfo{publisher}{Springer}, \bibinfo{year}{2017}).

\bibitem{sdcspatial2019}
\bibinfo{author}{{de Jonge}, E.} \& \bibinfo{author}{{de Wolf}, P.-P.}
\newblock \textit{\bibinfo{title}{sdcSpatial: Statistical Disclosure Control
  for Spatial Data}} (\bibinfo{year}{2019}).
\newblock \urlprefix\url{https://CRAN.R-project.org/package=sdcSpatial}.
\newblock \bibinfo{note}{R package version 0.1.1}.

\bibitem{DHS2013}
\bibinfo{author}{Burgert, C.~R.}, \bibinfo{author}{Colston, J.},
  \bibinfo{author}{Roy, T.} \& \bibinfo{author}{Zachary, B.}
\newblock \bibinfo{title}{{Geographic Displacement Procedure and Georeferenced
  Data Release Health Surveys. DHS Spatial Analysis Reports}}.
\newblock \bibinfo{type}{Tech. Rep.} \bibinfo{number}{7},
  \bibinfo{institution}{{ICF International, USAID}},
  \bibinfo{address}{Calverton, Maryland, USA} (\bibinfo{year}{2013}).
\newblock \urlprefix\url{https://dhsprogram.com/pubs/pdf/SAR7/SAR7.pdf}.

\bibitem{Rocher2019EstimatingModels}
\bibinfo{author}{Rocher, L.}, \bibinfo{author}{Hendrickx, J.~M.} \&
  \bibinfo{author}{de~Montjoye, Y.~A.}
\newblock \bibinfo{title}{{Estimating the success of re-identifications in
  incomplete datasets using generative models}}.
\newblock \textit{\bibinfo{journal}{Nat. Commun.}}
  \textbf{\bibinfo{volume}{10}}, \bibinfo{pages}{3069} (\bibinfo{year}{2019}).
\newblock \urlprefix\url{https://doi.org/10.1038/s41467-019-10933-3}.

\bibitem{elkies2015scrambling}
\bibinfo{author}{Elkies, N.}, \bibinfo{author}{Fink, G.} \&
  \bibinfo{author}{B{\"a}rnighausen, T.}
\newblock \bibinfo{title}{{“Scrambling” geo-referenced data to protect
  privacy induces bias in distance estimation}}.
\newblock \textit{\bibinfo{journal}{Popul. Environ.}}
  \textbf{\bibinfo{volume}{37}}, \bibinfo{pages}{83--98}
  (\bibinfo{year}{2015}).
\newblock \urlprefix\url{https://doi.org/10.1007/s11111-014-0225-0}.

\bibitem{warren2016influence}
\bibinfo{author}{Warren, J.~L.}, \bibinfo{author}{Perez-Heydrich, C.},
  \bibinfo{author}{Burgert, C.~R.} \& \bibinfo{author}{Emch, M.~E.}
\newblock \bibinfo{title}{Influence of demographic and health survey point
  displacements on distance-based analyses}.
\newblock \textit{\bibinfo{journal}{Spatial Demography}}
  \textbf{\bibinfo{volume}{4}}, \bibinfo{pages}{155--173}
  (\bibinfo{year}{2016}).
\newblock \urlprefix\url{https://doi.org/10.1007/s40980-015-0014-0}.

\bibitem{Blankespoor2021}
\bibinfo{author}{Blankespoor, B.}, \bibinfo{author}{Croft, T.},
  \bibinfo{author}{Dontamsetti, T.}, \bibinfo{author}{Mayala, B.} \&
  \bibinfo{author}{Murray, S.}
\newblock \bibinfo{title}{{Spatial anonymization: Guidance note prepared for
  the Inter-Secretariat working group on household surveys}}.
\newblock \bibinfo{type}{Tech. Rep.}, \bibinfo{institution}{UN
  Inter-secretariat Working Group on Household Surveys Task Force on Spatial
  Anonymization in Public-Use Household Survey Datasets}
  (\bibinfo{year}{2021}).
\newblock
  \urlprefix\url{https://unstats.un.org/iswghs/task-forces/documents/Spatial_Anonymization_Report_submit01272021_ISWGHS.pdf}.

\bibitem{hunter2021working}
\bibinfo{author}{Hunter, L.~M.} \textit{et~al.}
\newblock \bibinfo{title}{{Working toward effective anonymization for
  surveillance data: innovation at South Africa’s Agincourt Health and
  Socio-Demographic Surveillance Site}}.
\newblock \textit{\bibinfo{journal}{Popul. Environ.}}
  \textbf{\bibinfo{volume}{42}}, \bibinfo{pages}{445--476}
  (\bibinfo{year}{2021}).
\newblock \urlprefix\url{https://doi.org/10.1007/s11111-020-00372-4}.

\bibitem{drechsler2008new}
\bibinfo{author}{Drechsler, J.}, \bibinfo{author}{Dundler, A.},
  \bibinfo{author}{Bender, S.}, \bibinfo{author}{R{\"a}ssler, S.} \&
  \bibinfo{author}{Zwick, T.}
\newblock \bibinfo{title}{A new approach for disclosure control in the iab
  establishment panel—multiple imputation for a better data access}.
\newblock \textit{\bibinfo{journal}{Adv. Stat. Anal.}}
  \textbf{\bibinfo{volume}{92}}, \bibinfo{pages}{439--458}
  (\bibinfo{year}{2008}).
\newblock \urlprefix\url{https://doi.org/10.1007/s10182-008-0090-1}.

\bibitem{heldal2019synthetic}
\bibinfo{author}{Heldal, J.} \& \bibinfo{author}{Iancu, D.-C.}
\newblock \bibinfo{title}{{Synthetic data generation for anonymization
  purposes. Application on the Norwegian Survey on living conditions/EHIS}}.
\newblock In \textit{\bibinfo{booktitle}{{Joint UNECE/Eurostat Work Session on
  Statistical Data Confidentiality}}} (\bibinfo{year}{2019}).
\newblock
  \urlprefix\url{https://unece.org/fileadmin/DAM/stats/documents/ece/ces/ge.46/2019/mtg1/SDC2019_S1_Norway_Heldal_Iancu_AD.pdf}.

\bibitem{alfons2011simulation}
\bibinfo{author}{Alfons, A.}, \bibinfo{author}{Kraft, S.},
  \bibinfo{author}{Templ, M.} \& \bibinfo{author}{Filzmoser, P.}
\newblock \bibinfo{title}{{Simulation of close-to-reality population data for
  household surveys with application to EU-SILC}}.
\newblock \textit{\bibinfo{journal}{Stat. Methods Appt.}}
  \textbf{\bibinfo{volume}{20}}, \bibinfo{pages}{383--407}
  (\bibinfo{year}{2011}).
\newblock \urlprefix\url{https://doi.org/10.1007/s10260-011-0163-2}.

\bibitem{Templ2017simPop}
\bibinfo{author}{Templ, M.}, \bibinfo{author}{Meindl, B.},
  \bibinfo{author}{Kowarik, A.} \& \bibinfo{author}{Dupriez, O.}
\newblock \bibinfo{title}{{Simulation of synthetic complex data: The R package
  simPop}}.
\newblock \textit{\bibinfo{journal}{J. Stat. Softw.}}
  \textbf{\bibinfo{volume}{79}}, \bibinfo{pages}{1--38} (\bibinfo{year}{2017}).
\newblock
  \urlprefix\url{https://www.jstatsoft.org/index.php/jss/article/view/v079i10}.

\bibitem{reiter2005}
\bibinfo{author}{Reiter, J.~P.}
\newblock \bibinfo{title}{{Using CART to generate partially synthetic public
  use microdata}}.
\newblock \textit{\bibinfo{journal}{J. Off. Stat.}}
  \textbf{\bibinfo{volume}{21}}, \bibinfo{pages}{441--462}
  (\bibinfo{year}{2005}).
\newblock
  \urlprefix\url{https://www.scb.se/contentassets/ca21efb41fee47d293bbee5bf7be7fb3/using-cart-to-generate-partially-synthetic-public-use-microdata.pdf}.

\bibitem{wang2012multiple}
\bibinfo{author}{Wang, H.} \& \bibinfo{author}{Reiter, J.~P.}
\newblock \bibinfo{title}{Multiple imputation for sharing precise geographies
  in public use data}.
\newblock \textit{\bibinfo{journal}{Ann. Appl. Stat.}}
  \textbf{\bibinfo{volume}{6}}, \bibinfo{pages}{229--252}
  (\bibinfo{year}{2012}).
\newblock \urlprefix\url{https://doi.org/10.1214/11-AOAS506}.

\bibitem{Li2014}
\bibinfo{author}{Li, H.}, \bibinfo{author}{Xiong, L.} \&
  \bibinfo{author}{Jiang, X.}
\newblock \bibinfo{title}{Differentially private synthesization of
  multi-dimensional data using copula functions}.
\newblock In \textit{\bibinfo{booktitle}{Advances in database technology:
  proceedings. International conference on extending database technology}},
  \bibinfo{pages}{475--486} (\bibinfo{year}{2014}).
\newblock \urlprefix\url{https://doi.org/10.5441/002/edbt.2014.43}.

\bibitem{zhang2017}
\bibinfo{author}{Zhang, J.}, \bibinfo{author}{Cormode, G.},
  \bibinfo{author}{Procopiuc, C.~M.}, \bibinfo{author}{Srivastava, D.} \&
  \bibinfo{author}{Xiao, X.}
\newblock \bibinfo{title}{Privbayes: Private data release via bayesian
  networks}.
\newblock \textit{\bibinfo{journal}{ACM Trans. Database Syst.}}
  \textbf{\bibinfo{volume}{42}}, \bibinfo{pages}{1--41} (\bibinfo{year}{2017}).
\newblock \urlprefix\url{https://doi.org/10.1145/3134428}.

\bibitem{Sun2019}
\bibinfo{author}{Sun, Y.}, \bibinfo{author}{Cuesta-Infante, A.} \&
  \bibinfo{author}{Veeramachaneni, K.}
\newblock \bibinfo{title}{Learning vine copula models for synthetic data
  generation}.
\newblock \textit{\bibinfo{journal}{Proc. AAAI Conf. Artif. Intell.}}
  \textbf{\bibinfo{volume}{33}}, \bibinfo{pages}{5049--5057}
  (\bibinfo{year}{2019}).
\newblock \urlprefix\url{https://doi.org/10.1609/aaai.v33i01.33015049}.

\bibitem{torkzadehmahani2019}
\bibinfo{author}{Torkzadehmahani, R.}, \bibinfo{author}{Kairouz, P.} \&
  \bibinfo{author}{Paten, B.}
\newblock \bibinfo{title}{Dp-cgan: Differentially private synthetic data and
  label generation}.
\newblock In \textit{\bibinfo{booktitle}{Proceedings of the IEEE/CVF Conference
  on Computer Vision and Pattern Recognition Workshops}}
  (\bibinfo{year}{2019}).
\newblock
  \urlprefix\url{https://openaccess.thecvf.com/content_CVPRW_2019/html/CV-COPS/Torkzadehmahani_DP-CGAN_Differentially_Private_Synthetic_Data_and_Label_Generation_CVPRW_2019_paper.html}.

\bibitem{Xu2019}
\bibinfo{author}{Xu, L.}, \bibinfo{author}{Skoularidou, M.},
  \bibinfo{author}{Cuesta-Infante, A.} \& \bibinfo{author}{Veeramachaneni, K.}
\newblock \bibinfo{title}{{Modeling tabular data using conditional GAN}}.
\newblock \textit{\bibinfo{journal}{Adv. Neural Inf. Process. Syst.}}
  \textbf{\bibinfo{volume}{32}}, \bibinfo{pages}{1--11} (\bibinfo{year}{2019}).
\newblock
  \urlprefix\url{https://proceedings.neurips.cc/paper/2019/hash/254ed7d2de3b23ab10936522dd547b78-Abstract.html}.

\bibitem{dhswebsite}
\bibinfo{author}{{ICF}}.
\newblock \bibinfo{title}{{The DHS Program Spatial Data Repository.}}
  (\bibinfo{year}{2022}).
\newblock \urlprefix\url{https://spatialdata.dhsprogram.com/home/}.

\bibitem{WorldPopwww.worldpop.org-SchoolofGeographyandEnvironmentalScienceUniversityofSouthamptonDepartmentofGeographyandGeosciencesUniversityofLouisvilleDepartementdeGeographie2018WorldPopb1}
\bibinfo{author}{{WorldPop}}.
\newblock \bibinfo{title}{{Global High Resolution Population Denominators
  Project}} (\bibinfo{year}{2018}).
\newblock \urlprefix\url{www.worldpop.org}.

\bibitem{sklar1959}
\bibinfo{author}{Sklar, A.}
\newblock \bibinfo{title}{Fonctions de répartition à n dimensions et leurs
  marges}.
\newblock \textit{\bibinfo{journal}{Publ. Inst. Statist. Univ. Paris}}
  \textbf{\bibinfo{volume}{8}}, \bibinfo{pages}{229--231}
  (\bibinfo{year}{1959}).

\bibitem{kamthe2021copula}
\bibinfo{author}{Kamthe, S.}, \bibinfo{author}{Assefa, S.} \&
  \bibinfo{author}{Deisenroth, M.}
\newblock \bibinfo{title}{Copula flows for synthetic data generation}.
\newblock \textit{\bibinfo{journal}{arXiv preprint arXiv:2101.00598}}
  (\bibinfo{year}{2021}).

\bibitem{AMELI2011}
\bibinfo{author}{Alfons, A.} \textit{et~al.}
\newblock \bibinfo{title}{{Synthetic Data Generation of SILC Data. Research
  Project Report WP6, D6.2}}.
\newblock \bibinfo{type}{Tech. Rep.}, \bibinfo{institution}{{The AMELI
  Project}} (\bibinfo{year}{2011}).
\newblock
  \urlprefix\url{https://www.uni-trier.de/fileadmin/fb4/projekte/SurveyStatisticsNet/Ameli_Delivrables/AMELI-WP6-D6.2-240611.pdf}.

\bibitem{Koebe2022}
\bibinfo{author}{Koebe, T.}, \bibinfo{author}{Arias-Salazar, A.},
  \bibinfo{author}{Rojas-Perilla, N.} \& \bibinfo{author}{Schmid, T.}
\newblock \bibinfo{title}{{Intercensal updating using structure-preserving
  methods and satellite imagery}}.
\newblock \textit{\bibinfo{journal}{J. R. Stat. Soc. Ser. A Stat. Soc.}}
  (\bibinfo{year}{2022}).
\newblock
  \urlprefix\url{https://onlinelibrary.wiley.com/doi/full/10.1111/rssa.12802}.

\bibitem{sdv2022}
\bibinfo{author}{{MIT Data To AI Lab}}.
\newblock \bibinfo{title}{{The synthetic data vault (SDV)}}.
\newblock \bibinfo{howpublished}{\url{https://sdv.dev/}}
  (\bibinfo{year}{2022}).

\bibitem{patki2016synthetic}
\bibinfo{author}{{Patki}, N.}, \bibinfo{author}{{Wedge}, R.} \&
  \bibinfo{author}{{Veeramachaneni}, K.}
\newblock \bibinfo{title}{The synthetic data vault}.
\newblock In \textit{\bibinfo{booktitle}{2016 IEEE International Conference on
  Data Science and Advanced Analytics (DSAA)}}, \bibinfo{pages}{399--410}
  (\bibinfo{year}{2016}).

\bibitem{jeong2016copula}
\bibinfo{author}{Jeong, B.}, \bibinfo{author}{Lee, W.}, \bibinfo{author}{Kim,
  D.-S.} \& \bibinfo{author}{Shin, H.}
\newblock \bibinfo{title}{Copula-based approach to synthetic population
  generation}.
\newblock \textit{\bibinfo{journal}{PloS ONE}} \textbf{\bibinfo{volume}{11}},
  \bibinfo{pages}{e0159496} (\bibinfo{year}{2016}).
\newblock \urlprefix\url{https://doi.org/10.1371/journal.pone.0159496}.

\bibitem{janke2021}
\bibinfo{author}{Janke, T.}, \bibinfo{author}{Ghanmi, M.} \&
  \bibinfo{author}{Steinke, F.}
\newblock \bibinfo{title}{Implicit generative copulas}.
\newblock In \bibinfo{editor}{Ranzato, M.}, \bibinfo{editor}{Beygelzimer, A.},
  \bibinfo{editor}{Dauphin, Y.}, \bibinfo{editor}{Liang, P.} \&
  \bibinfo{editor}{Vaughan, J.~W.} (eds.) \textit{\bibinfo{booktitle}{Adv.
  Neural Inf. Process. Syst.}}, vol.~\bibinfo{volume}{34},
  \bibinfo{pages}{26028--26039} (\bibinfo{publisher}{Curran Associates, Inc.},
  \bibinfo{year}{2021}).
\newblock
  \urlprefix\url{https://proceedings.neurips.cc/paper/2021/file/dac4a67bdc4a800113b0f1ad67ed696f-Paper.pdf}.

\bibitem{nelsen2007introduction}
\bibinfo{author}{Nelsen, R.~B.}
\newblock \textit{\bibinfo{title}{An introduction to copulas}}
  (\bibinfo{publisher}{Springer Science \& Business Media},
  \bibinfo{year}{2007}).

\bibitem{Mendez2011}
\bibinfo{author}{M{\'{e}}ndez, F.} \& \bibinfo{author}{Bravo, O.}
\newblock \bibinfo{title}{{Costa Rica Mapas de Pobreza 2011}}.
\newblock \bibinfo{type}{Tech. Rep.}, \bibinfo{institution}{INEC Costa Rica},
  \bibinfo{address}{San Jos{\'{e}}, Costa Rica} (\bibinfo{year}{2011}).
\newblock
  \urlprefix\url{https://www.inec.cr/sites/default/files/documentos/pobreza_y_presupuesto_de_hogares/pobreza/metodologias/documentos_metodologicos/mepobrezacenso2011-01.pdf.pdf}.

\bibitem{Alkire2019The2019}
\bibinfo{author}{Alkire, S.}, \bibinfo{author}{Kanagaratnam, U.} \&
  \bibinfo{author}{Suppa, N.}
\newblock \bibinfo{title}{{The Global Multidimensional Poverty Index (MPI)
  2019. OPHI MPI Methodological Note 47}}.
\newblock \bibinfo{type}{Tech. Rep.}, \bibinfo{institution}{Oxford Poverty and
  Human Development Initiative, University of Oxford} (\bibinfo{year}{2019}).
\newblock
  \urlprefix\url{https://www.ophi.org.uk/wp-content/uploads/OPHI_MPI_MN_47_2019_vs2.pdf}.

\bibitem{Zhang2021}
\bibinfo{author}{Zhang, Z.} \textit{et~al.}
\newblock \bibinfo{title}{{PrivSyn: Differentially private data synthesis}}.
\newblock In \textit{\bibinfo{booktitle}{Proceedings of the 30th USENIX
  Security Symposium}}, \bibinfo{pages}{929--946} (\bibinfo{year}{2021}).
\newblock
  \urlprefix\url{https://www.usenix.org/system/files/sec21fall-zhang-zhikun.pdf}.
\newblock \eprint{2012.15128}.

\bibitem{Jordon2019}
\bibinfo{author}{Jordon, J.}, \bibinfo{author}{Yoon, J.} \&
  \bibinfo{author}{{Van Der Schaar}, M.}
\newblock \bibinfo{title}{{PATE-GaN: Generating synthetic data with
  differential privacy guarantees}}.
\newblock In \textit{\bibinfo{booktitle}{International Conference on Learning
  Representations}} (\bibinfo{year}{2019}).
\newblock \urlprefix\url{https://openreview.net/forum?id=S1zk9iRqF7}.

\bibitem{Benali2021}
\bibinfo{author}{Benali, F.}, \bibinfo{author}{Bod{\'e}n{\`e}s, D.},
  \bibinfo{author}{Labroche, N.} \& \bibinfo{author}{de~Runz, C.}
\newblock \bibinfo{title}{{MTCopula: Synthetic complex data generation using
  copula}}.
\newblock In \textit{\bibinfo{booktitle}{23rd International Workshop on Design,
  Optimization, Languages and Analytical Processing of Big Data (DOLAP)}},
  \bibinfo{pages}{51--60} (\bibinfo{year}{2021}).
\newblock \urlprefix\url{https://hal.archives-ouvertes.fr/hal-03188317}.

\bibitem{Bourou2021}
\bibinfo{author}{Bourou, S.}, \bibinfo{author}{{El Saer}, A.},
  \bibinfo{author}{Velivassaki, T.~H.}, \bibinfo{author}{Voulkidis, A.} \&
  \bibinfo{author}{Zahariadis, T.}
\newblock \bibinfo{title}{{A Review of Tabular Data Synthesis Using GANs on an
  IDS Dataset}}.
\newblock \textit{\bibinfo{journal}{Information}}
  \textbf{\bibinfo{volume}{12}}, \bibinfo{pages}{375--389}
  (\bibinfo{year}{2021}).
\newblock \urlprefix\url{https://www.mdpi.com/2078-2489/12/9/375}.

\bibitem{jiangdan2020}
\bibinfo{author}{Jiang, D.}, \bibinfo{author}{Lin, W.} \&
  \bibinfo{author}{Raghavan, N.}
\newblock \bibinfo{title}{A novel framework for semiconductor manufacturing
  final test yield classification using machine learning techniques}.
\newblock \textit{\bibinfo{journal}{IEEE Access}} \textbf{\bibinfo{volume}{8}},
  \bibinfo{pages}{197885--197895} (\bibinfo{year}{2020}).

\bibitem{mansfield1977medical}
\bibinfo{author}{Mansfield, P.} \& \bibinfo{author}{Maudsley, A.~A.}
\newblock \bibinfo{title}{Medical imaging by nmr}.
\newblock \textit{\bibinfo{journal}{Br. J. Radiol.}}
  \textbf{\bibinfo{volume}{50}}, \bibinfo{pages}{188--194}
  (\bibinfo{year}{1977}).

\bibitem{aizawa2003information}
\bibinfo{author}{Aizawa, A.}
\newblock \bibinfo{title}{An information-theoretic perspective of tf--idf
  measures}.
\newblock \textit{\bibinfo{journal}{Inf. Process. Manag.}}
  \textbf{\bibinfo{volume}{39}}, \bibinfo{pages}{45--65}
  (\bibinfo{year}{2003}).

\bibitem{Sabharwal2021}
\bibinfo{author}{Sabharwal, N.} \& \bibinfo{author}{Agrawal, A.}
\newblock \textit{\bibinfo{title}{Introduction to Word Embeddings}},
  \bibinfo{pages}{41—63} (\bibinfo{publisher}{Apress},
  \bibinfo{address}{Berkeley, CA}, \bibinfo{year}{2021}).
\newblock \urlprefix\url{https://doi.org/10.1007/978-1-4842-6664-9_3}.

\bibitem{fay1979estimates}
\bibinfo{author}{Fay, R.~E.} \& \bibinfo{author}{Herriot, R.~A.}
\newblock \bibinfo{title}{{Estimates of income for small places: an application
  of James-Stein procedures to census data}}.
\newblock \textit{\bibinfo{journal}{J. Am. Stat. Assoc.}}
  \textbf{\bibinfo{volume}{74}}, \bibinfo{pages}{269--277}
  (\bibinfo{year}{1979}).
\newblock \urlprefix\url{https://doi.org/10.1080/01621459.1979.10482505}.

\bibitem{RaoMol15}
\bibinfo{author}{Rao, J. N.~K.} \& \bibinfo{author}{Molina, I.}
\newblock \textit{\bibinfo{title}{{Small Area Estimation}}}
  (\bibinfo{publisher}{Wiley}, \bibinfo{address}{New York},
  \bibinfo{year}{2015}), \bibinfo{edition}{2nd} edn.

\bibitem{Rica}
\bibinfo{author}{{Instituto Nacional de Estadistica y Censos}}.
\newblock \bibinfo{title}{{X Censo Nacional de Poblaci{\'{o}}n y VI de
  Vivienda. Cat{\'{a}}logo central de datos}} (\bibinfo{year}{2022}).
\newblock \urlprefix\url{http://sistemas.inec.cr/pad5/index.php/catalog/113}.

\end{thebibliography}

\newpage

% \section*{Author Contributions}
% T.K. and A.A.S. designed and performed experiments and wrote the paper. T.S. designed the experiments.

% \section*{Competing interests}
% The authors declare no competing interests.

% \section*{Data Availability}
% The 10\% sample of the Costa Rican census dataset is available from the microdata catalogue of the national statistical office of Costa Rica -- INEC -- under a licensing agreement at:\\ \hyperlink{http://sistemas.inec.cr/pad5/index.php/catalog/113}{http://sistemas.inec.cr/pad5/index.php/catalog/113}.

% \noindent The satellite-derived covariates are openly available from the WorldPop repository at:\\ \hyperlink{https://www.worldpop.org/doi/10.5258/SOTON/WP00644}{https://www.worldpop.org/doi/10.5258/SOTON/WP00644}.

% \section*{Code Availability}
% All simulations were implemented in Python (Python 3.7.3) and R (R version 4.2.0) on a standard 64-bit machine with Kubuntu 20.04, 8 × Intel Core i7-8550U CPU @ 1.80GHz and 23,4 GiB of RAM. All code necessary to replicate the findings is available at: \hyperlink{https://github.com/tilluz/survey\_releases}{https://github.com/tilluz/survey\_releases}.

\section{Supplementary information}

\subsection{Data description}

As our reference dataset in this project, we use data from Costa Rica -- notably the X\textsuperscript{th} Population and VI\textsuperscript{th} Housing Census of Costa Rica, 2011 (Censo Nacional de población y Viviendas de Costa Rica 2011) -- to produce three different data file types: First, we draw survey samples from a census population using a stratified two-stage cluster sample design without applying any statistical disclosure control mechanisms. We use these survey samples (called \textit{true} surveys in the study) as starting point for creating file types two and three: By re-assigning clusters to new zip codes based on the displacement algorithm described in Algorithm 1 %\ref{alg:one}
, we perturb the zip code identifier in the true surveys, thereby creating the \textit{geomasked} surveys. Again based on the true surveys, we apply the copula-based synthetic data generation algorithm described in Algorithm 2 %\ref{alg:two}
to generate synthetic data for each attribute except the zip code, which keeps it original structure. In addition, in order to test the robustness of our specifications, we create additional datasets with alternating data generating process designs. The censuses are carried out every ten years by the national statistic office of Costa Rica (INEC) and collect information of people, households, and dwellings on topics such as access to education, employment, social security, technology necessary for the planning, execution, and evaluation of public policies \citep{Mendez2011}.

Administratively, Costa Rica had in 2011 four disaggregation levels: two zones, six planning regions, 81 cantons and 473 districts (municipalities). The sampling design used for the main National Household Survey (Encuesta Nacional de Hogares, ENAHO) specifies twelve strata - each planning region divided by urban and rural areas. In this case, the strata coincide with the study domains. %In this census, 10,461 primary sampling units (PSUs) and 1,359,168 dwellings were identified. 
% The sampling design used in the ENAHO is a two-stage stratified random sampling where census segments are the first stage units selected with probability proportional to size, and dwellings are defined as the final stage units. %Smaller domains are not considered to guarantee a coefficient of variation less than 15\% for the main poverty measure (percentage of household under poverty)
For our experiment, we use a 10\% random sample of the original 2011 census, which can be obtained from \cite{Rica} as a pseudo-population. The smallest geographical information available in this dataset are the 473 districts. In the first stage, we select districts as our PSUs for each stratum separately with a selection probability proportional to population size. In the second stage, we select a minimum of 10 households in each PSU by using simple random sampling without replacement. PSUs with less than 10 households are discarded from this procedure, affecting roughly 4\% of all PSUs.

\begin{table}[ht]
\centering
\small{
    \begin{tabular}{ccccc}
      \hline
     $N_C$ & $n_D$ & \# of PSUs in $C$ & \# of PSUs in $D$ & \# of attributes \\ 
      \hline
    427830 & [7638; 11914] & 767 & 123 & 106 \\
       \hline
    \end{tabular}
	\caption{\textbf{Descriptive statistics on the census-derived data across 100 simulation runs}}
	\label{tab:appendix-descriptives}
	}
\end{table}

As auxiliary information, we use covariates derived from satellite imagery. Specifically, we use features derived from satellite imagery provided by \cite{WorldPopwww.worldpop.org-SchoolofGeographyandEnvironmentalScienceUniversityofSouthamptonDepartmentofGeographyandGeosciencesUniversityofLouisvilleDepartementdeGeographie2018WorldPopb1} in our survey augmentation setup. The advantages of using satellite imagery here are five-fold: Data with virtually global coverage at high spatial resolutions for frequent time intervals on human-made impact provided in a structured format enables us to extract covariates for all administrative areas in Costa Rica at the time of the census. Therefore, we can use area-level survey augmentation (cf. Methods Section) %\ref{section:methods-fh}
to provide estimates, especially for areas not covered by the respective survey. WorldPop data are provided in the tagged image file format (TIFF) with a pixel representing roughly a 100m $\times$ 100m grid square in an open data repository under CC4.0 licence (\cite{WorldPopwww.worldpop.org-SchoolofGeographyandEnvironmentalScienceUniversityofSouthamptonDepartmentofGeographyandGeosciencesUniversityofLouisvilleDepartementdeGeographie2018WorldPopb1}). Pixel values are aggregated to the administrative areas of Costa Rica via their centroids. Specifically, we generate area-level averages for the distances to different types of natural areas (e.g. cultivated, woody-tree, and shrub areas, coastlines etc.) and to infrastructure such as roads and waterways, the intensity of night-time lights, topographic information and information on the presence of human settlements.

\subsection{Sensitivity of copula vis-à-vis geographic fitting level}

In order to study the effect of the geographic level on the copula modelling performed for synthetic data generation, we run Algorithm 2 %\ref{alg:two}
on the whole survey (`Country), the twelve strata (`Strata') and the roughly 110 zip code areas (`Zip Code'), respectively. Results are provided in Supplementary Fig. \ref{fig:appendix-sens-geo}. It appears that fitting the copula model on the whole survey limits the ability of the approach to capture regional variations. On the other hand, model fitting on the zip code-level does neither increase the re-identification risk of the zip code identifier as a private attribute and nor affect the overall prediction performance of the outcome variable, hinting at overfitting not being a problem on that level. Striking a balance between underlying sample size and a certain level of disaggregation shows better results. Also, it allows to scale computations to settings with larger samples and more attributes.

\begin{figure}[ht!]
\captionsetup[subfigure]{justification=centering}
     \centering
     \begin{subfigure}[c]{0.25\linewidth}
         \centering
         \includegraphics[width=\textwidth]{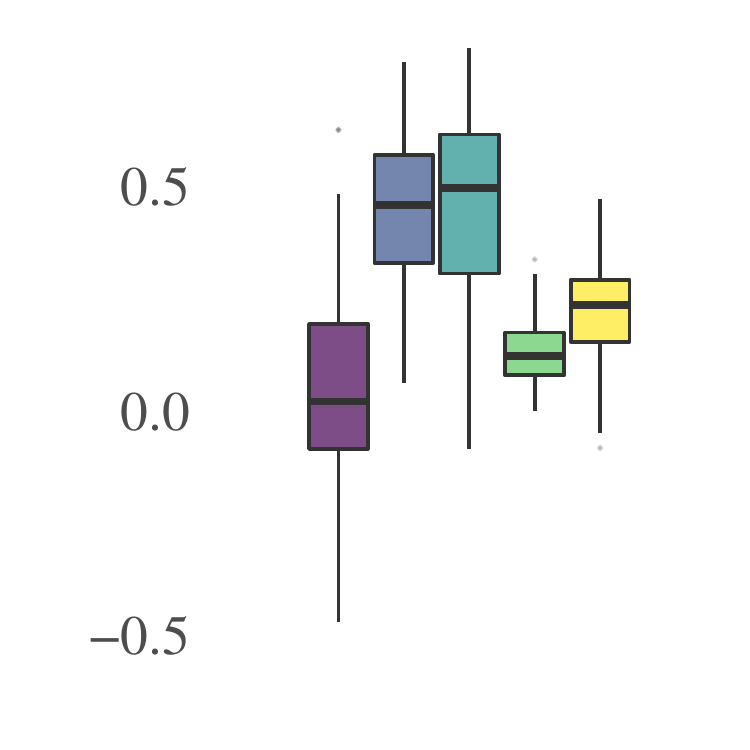}
         \caption{Adjusted $R^2$}
         \label{fig:appendix-pred-r2}
     \end{subfigure}
     \begin{subfigure}[c]{0.25\linewidth}
         \centering
         \includegraphics[width=\textwidth]{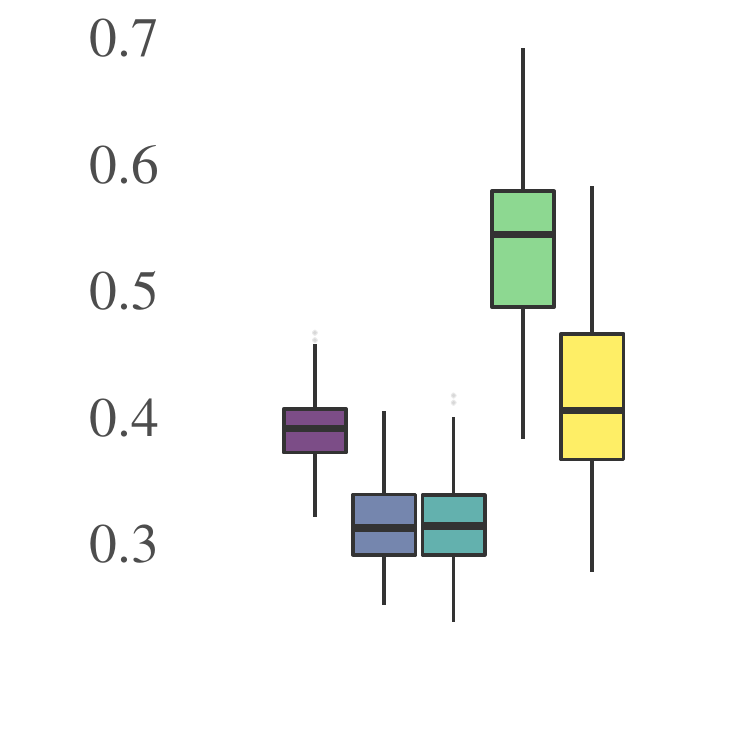}
         \caption{Relative Bias}
         \label{fig:appendix-pred-bias}
     \end{subfigure}
     \begin{subfigure}[c]{0.25\linewidth}
         \centering
         \includegraphics[width=\textwidth]{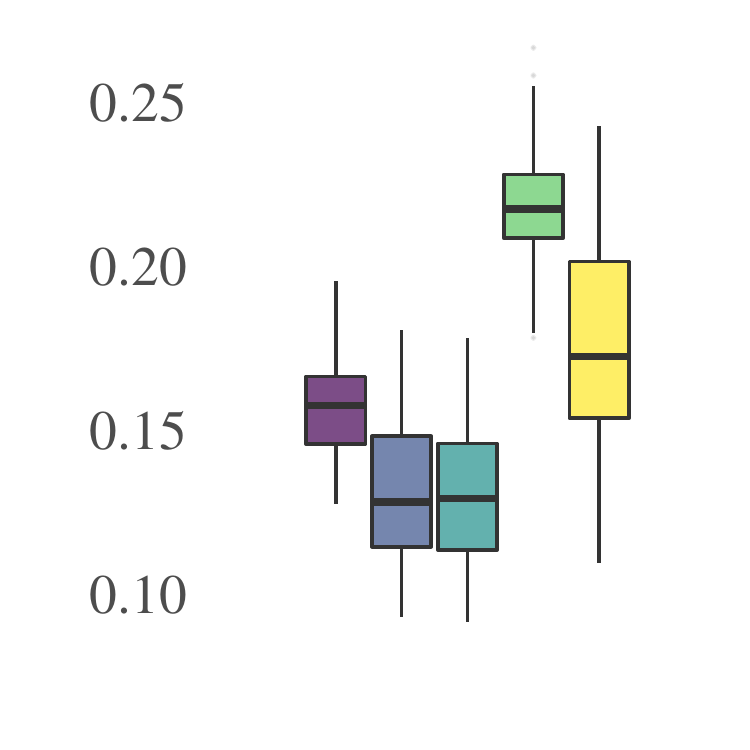}
         \caption{MSE}
         \label{fig:appendix-pred-rmse}
     \end{subfigure}
     \begin{subfigure}[c]{0.15\linewidth}
         \centering
         \includegraphics[width=\textwidth]{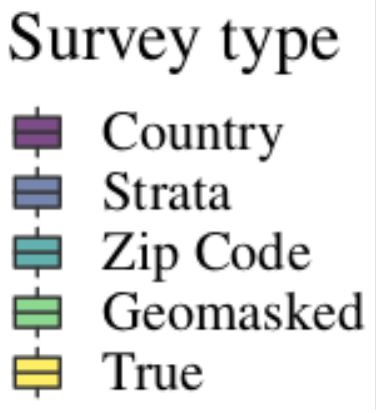}
        %  \caption{Synthetic survey}
         \label{fig:appendix-pred-legend}
     \end{subfigure}
     \hfill
     \begin{subfigure}[b]{0.49\linewidth}
         \centering
         \includegraphics[width=\textwidth]{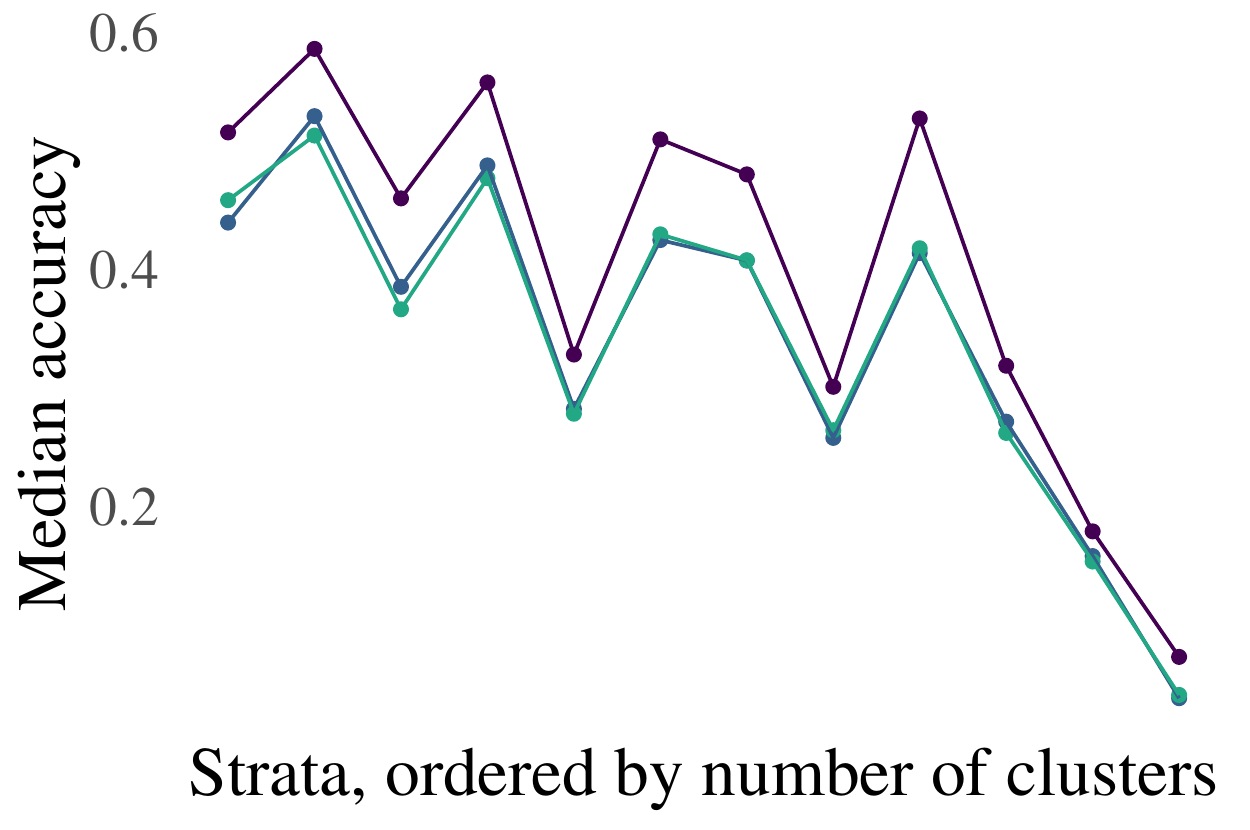}
         \caption{Re-identification Risk}
         \label{fig:appendix-risk}
     \end{subfigure}
     \begin{subfigure}[b]{0.49\linewidth}
         \centering
         \includegraphics[width=\textwidth]{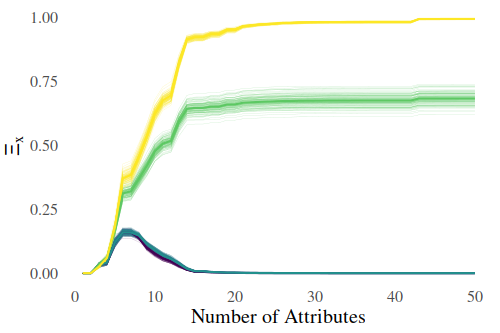}
         \caption{Population uniqueness across attributes}
         \label{fig:appendix-uniqueness}
     \end{subfigure}
        \caption{\textbf{Evaluation metrics for different geographical copula fitting levels.}\\ (a) - (c) The copula model is fitted on the whole survey (`Country'), for each of the twelve strata (`Strata') and for each of the roughly 110 zip codes (`Zip Code') separately. As a reference, the metrics for the geomasked and the true survey are provided as well. (d) The accuracy to successfully re-identify the zip code as a private attribute in the original data using a random forest model trained on synthetic data across fitting levels remains similar. (e) The share of population-unique survey respondents is virtually not affected by the copula fitting level.}
        \label{fig:appendix-sens-geo}
\end{figure}

% \begin{figure}
%     \captionsetup[subfigure]{justification=centering}
%     \centering
%     \begin{subfigure}[b]{\textwidth}
%     \centering
%     \begin{minipage}{0.85\textwidth}
%         \centering
%         \includegraphics[width=\textwidth]{viz/appendix_utility_performance_Adjusted_R2_GC_sample_100_1.pdf} % first figure itself
%         \caption{Adjusted $R^2$}
%         \centering
%         \includegraphics[width=\textwidth]{viz/appendix_utility_performance_Relative_Bias_GC_sample_100_1.pdf} % first figure itself
%         \caption{Relative Bias}
%         \centering
%         \includegraphics[width=\textwidth]{viz/appendix_utility_performance_RMSE_GC_sample_100_1.pdf}
%         \caption{RMSE}
%     \end{minipage}\hfill
%     \begin{minipage}{0.15\textwidth}
%         \centering
%         \includegraphics[width=\textwidth]{viz/appendix_utility_performance_legend_GC_sample_100_1.pdf} % second figure itself
%         % \caption{second figure}
%     \end{minipage}
%     \end{subfigure}
%     \caption{\textbf{Census NBI vs. survey-based NBI estimates.}\\
%         for one simulation run.}
%         \label{fig:appendix-pred}
% \end{figure}

\subsection{Effects of synthetic sample size on prediction outcomes}

\begin{figure}[ht!]
    \centering
    \includegraphics[width=\textwidth]{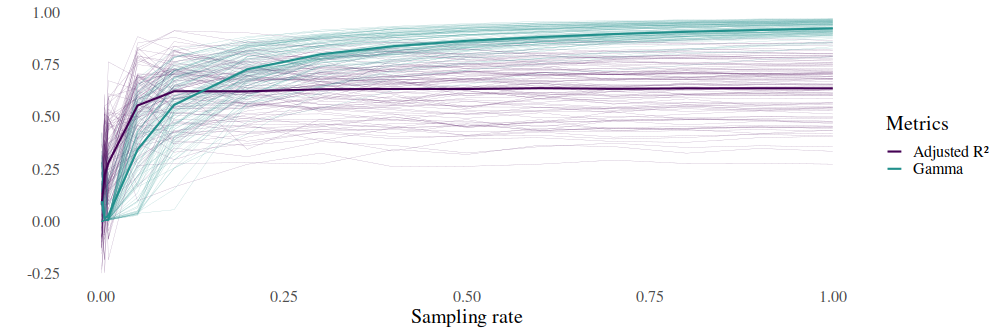}
    \caption{\textbf{Sensitivity of model performance on changes in synthetic sample size.}\\
    Samples are drawn from a synthetic population. The synthetic population is generated using the copula-based approach described in the Results Section. %\ref{section:results-copula}
    Sample sizes are determined by the sampling rate (shown on the x-axis). Results are evaluated against the true census population. The shrinkage factor $\gamma$ is averaged across zip codes. The thick lines represent the metric averages across the 100 simulation runs, the thin lines individual simulation runs.}
    \label{fig:appendix-gamma}
\end{figure}

Generative models can be used to create synthetic samples of an arbitrary size regardless the amount of underlying data. While the advantages of that are similar to those of other resampling procedures such as bootstrapping (i.e. to estimate the precision of the sample statistics or to perform cross-validation), it can also mislead modelling approaches that `borrow strength' from auxiliary data by overestimating the strength of the synthetic direct estimates eventually resulting in losses of explanatory power of the model. In our survey augmentation setup, the shrinkage factor $\gamma$ indicates whether final estimates rather rely on the direct estimates from the synthetic survey or on the satellite-derived covariates for the in-sample predictions depending on the sampling variance. Supplementary Figure \ref{fig:appendix-gamma} shows that larger sample sizes lead to increasing gamma values (via decreasing sampling variances of the direct estimator), however, not incurring losses in the goodness-of-fit of our estimation model. This hints at the fact that the contribution of the auxiliary information to the explanatory power of the model for the in-sample predictions is negligible.

\subsection{Choosing marginal distributions \& encoding schemes}

As already mentioned in the Section \ref{section:results-copula}, assuming normally-distributed margins may represent a misspecification of the true univariate distribution of $X_d$. In addition, computationally tractable alternatives to one-hot encoding exist. We compare two different ways to model the marginal distributions together with two different encoding schemes. The results are presented in Supplementary Fig. \ref{fig:appendix-encoding-marginals}. Measured by the normalized KL divergence averaged across 100 simulation runs, frequency encoding produces slightly better goodness-of-fit of the synthetic data ($Z_{KL} = 0.74$ for frequency encoding versus $Z_{KL} = 0.72$ for ordinal encoding with gaussian marginals). Surprisingly, the na\"ive assumption of normally distributed marginals outperforms the KS-based parametric marginals with $Z_{KL} = 0.74$ and $Z_{KL} = 0.70$, respectively.

\begin{figure}[ht!]
    \centering
    \includegraphics[width=\textwidth]{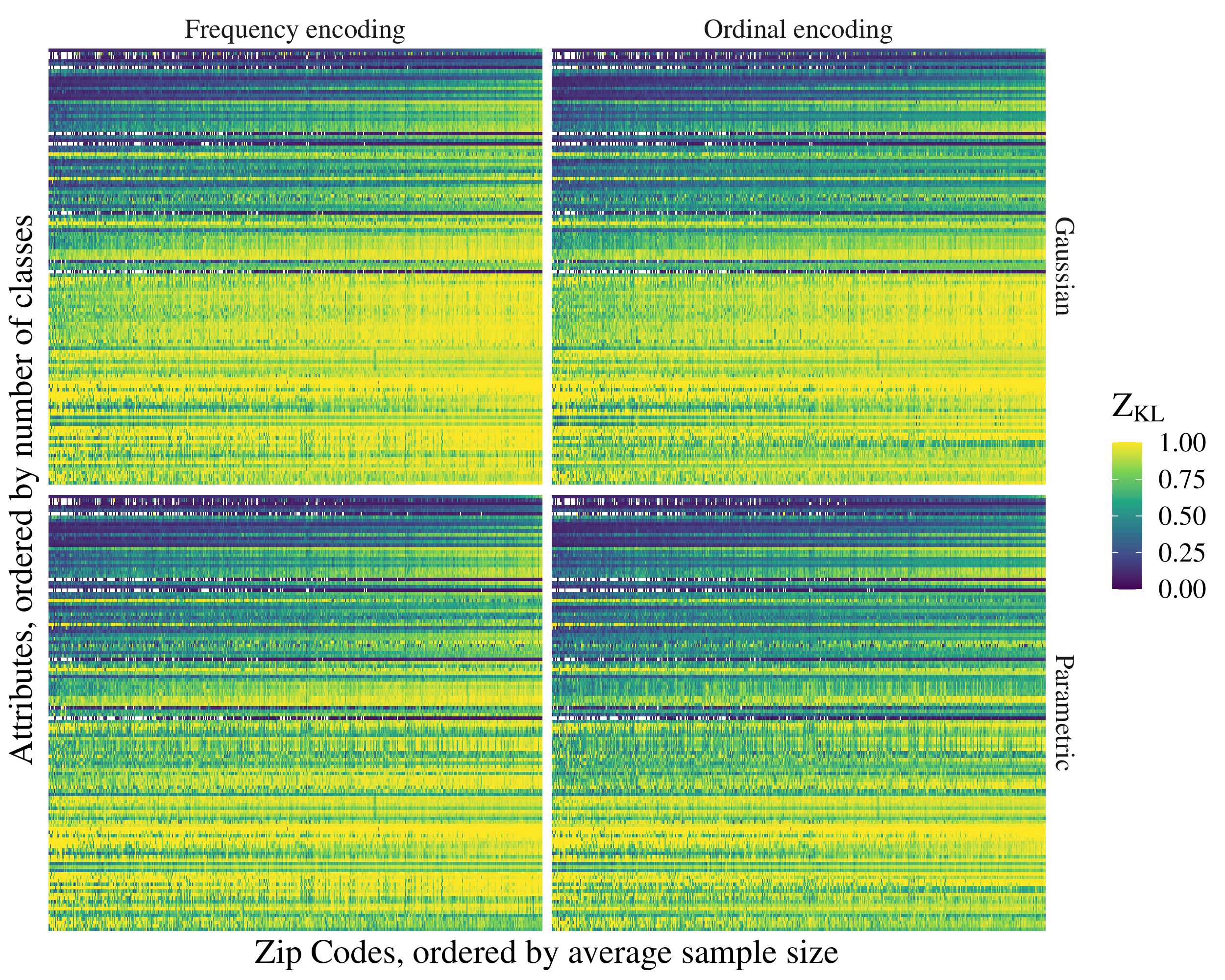}
    \caption{\textbf{Effect of encoding schemes and marginal distribution choice on the overall goodness-of-fit of the synthetic data measured by the normalized KL divergence $Z_{KL}$ (in bits).}\\
    The attributes on the y-axis are ordered by their respective number of classes, the zip codes on the x-axis are ordered by their average sample size across simulation rounds. Values close to one (yellow) represent little divergence from the true census distribution and therefore indicate a high goodness-of-fit.}
    \label{fig:appendix-encoding-marginals}
\end{figure}

\subsection{Detailed analysis of the NBI as composite indicator}

The NBI is a composite indicator computed from approx. 20 underlying survey variables grouped into four dimensions (i.e. access to decent housing (\textit{Acceso albergue digno}), access to a healthy life (\textit{Acceso a vida saludable}), access to knowledge (\textit{Acceso al conocimiento}) and access to other goods and services (\textit{Acceso a otros bienes y servicios})) using 19 indicators in total. All indicators and dimensions are binary (yes/no). An identified need in one of the indicators leads to a positive needs status in higher dimensions. The sensitivity for false positives is thus assumed to be high for the NBI as a small change (e.g. one year age difference) in one of the 19 underlying variables can turn a NBI-negative to a NBI-positive survey respondent.

Generally, two strategies for computed indicators exist to create synthetic counterparts: a) directly synthesize the computed indicators or b) re-construct the indicator based on synthetic survey variables. While the former is more likely to reflect the original distribution, it may not be consistently decomposable into its underlying indicators; vice-versa holds for the latter. The strength of these effects are largely determined by the complexity and sensitivity of the composite indicator and the overall goodness-of-fit of the synthetic data. Thus, if both approaches produce similar compositions, it can be regarded as a strong indication that the underlying synthetic data also successfully captures relationships across multiple variables in the dataset, not only the composite index. Supplementary Table \ref{table:appendix-pred} shows that this not fully holds for the NBI. 

\begin{table}[ht]
\centering
\begin{tabular}{lcccc}
  \hline
 Indicators & \# of indicators & Pearson's $\rho$ & $Z_{KL}$ & Incidence \\ 
  \hline
1.x & 5 & 0.42 & 0.99 & 100 \\ 
Dimension 1 &  & 0.24 & 0.98 & 647 \\ 
2.x & 5 & 0.22 & 0.98 & 85 \\ 
Dimension 2 &  & 0.19 & 0.98 & 455 \\ 
3.x & 2 & 0.02 & 0.89 & 507 \\ 
Dimension 3 &  & 0.02 & 0.84 & 1845 \\ 
4.x & 7 & 0.02 & 0.99 & 60 \\ 
Dimension 4 &  & 0.03 & 1.00 & 622 \\
\hline
Composite NBI & 19 & 0.07 & 0.97 & 3253 \\ 
   \hline
\end{tabular}
\caption{\textbf{Relationship between synthetic and computed NBI indicators across 100 simulation runs.}\\ Indicator-level results (e.g. 1.x) are averaged across indicators. The incidence describes the average number of respondents across 100 simulated surveys with unsatisfied needs in the respective indicator/dimension.}
\label{table:appendix-pred}
\end{table}

\begin{figure}[ht!]
\captionsetup[subfigure]{justification=centering}
     \centering
     \begin{subfigure}[c]{0.24\linewidth}
         \centering
         \includegraphics[width=\textwidth]{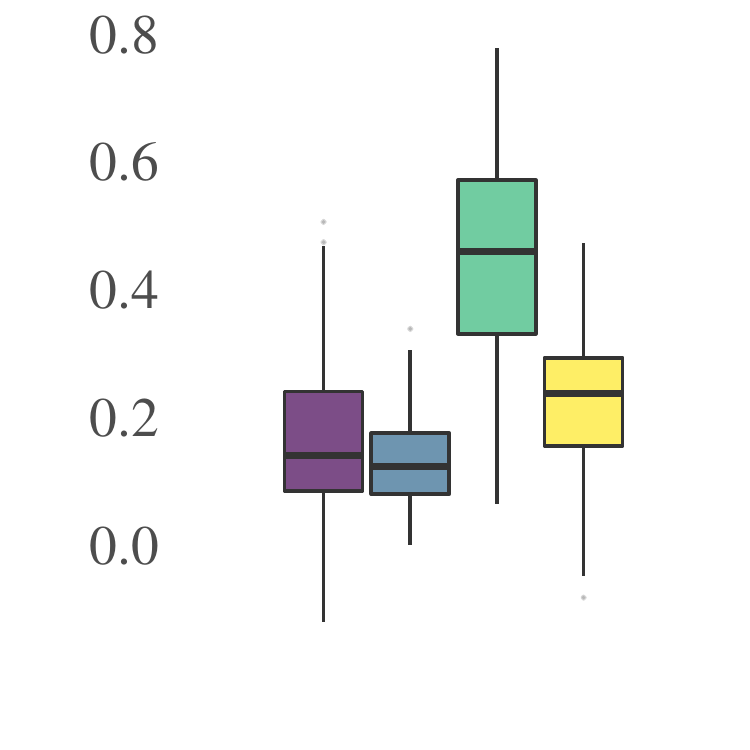}
         \caption{Adjusted R2}
         \label{fig:appendix-nbi-adj}
     \end{subfigure}
     \begin{subfigure}[c]{0.24\linewidth}
         \centering
         \includegraphics[width=\textwidth]{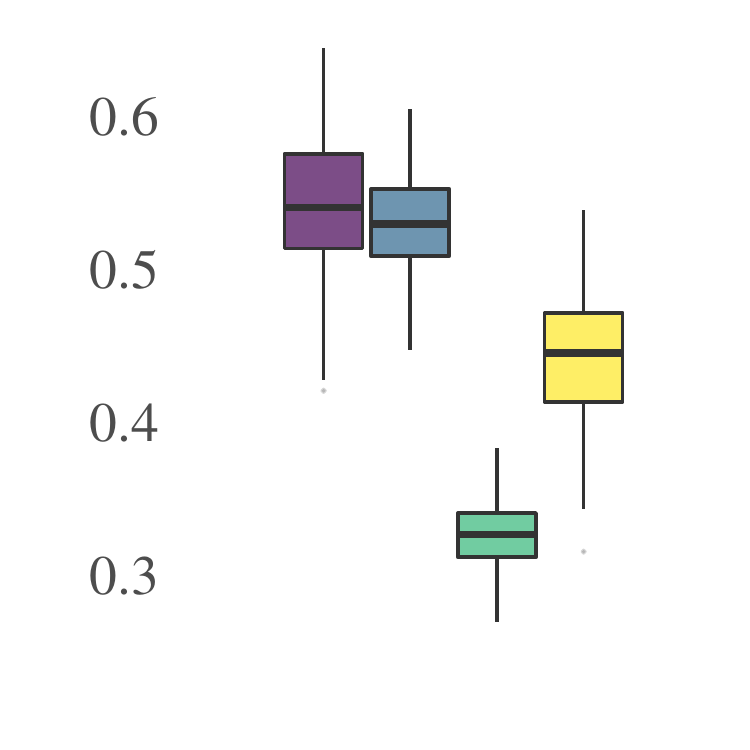}
         \caption{Relative Bias}
         \label{fig:appendix-nbi-bias}
     \end{subfigure}
     \begin{subfigure}[c]{0.24\linewidth}
         \centering
         \includegraphics[width=\textwidth]{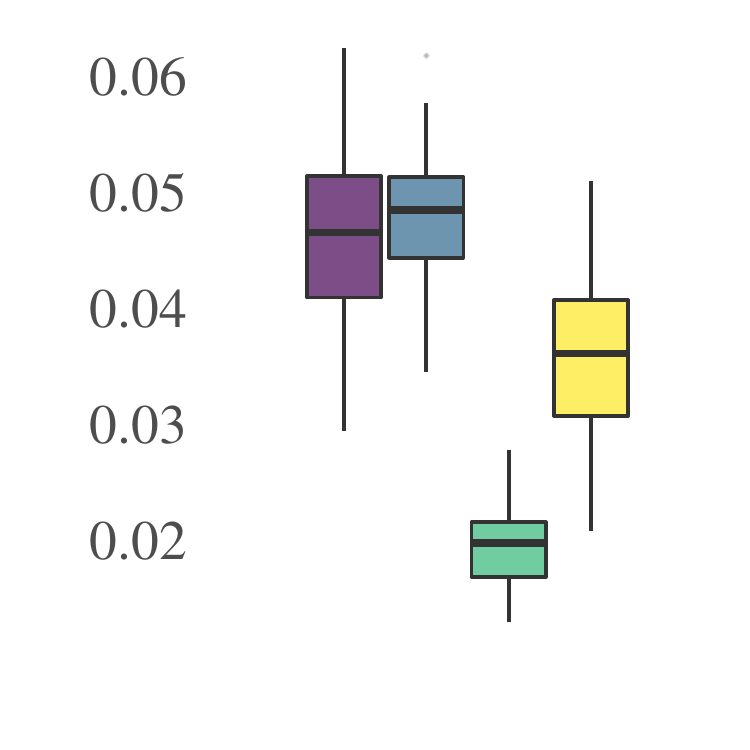}
         \caption{MSE}
         \label{fig:appendix-nbi-mse}
     \end{subfigure}
     \begin{subfigure}[b]{0.15\linewidth}
         \centering
         \includegraphics[width=\textwidth]{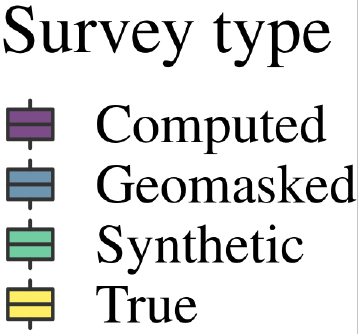}
        %  \caption{MSE}
        %  \label{fig:appendix-nbi-legend}
     \end{subfigure}
     \hfill
     \begin{subfigure}[b]{\linewidth}
         \centering
         \includegraphics[width=\textwidth]{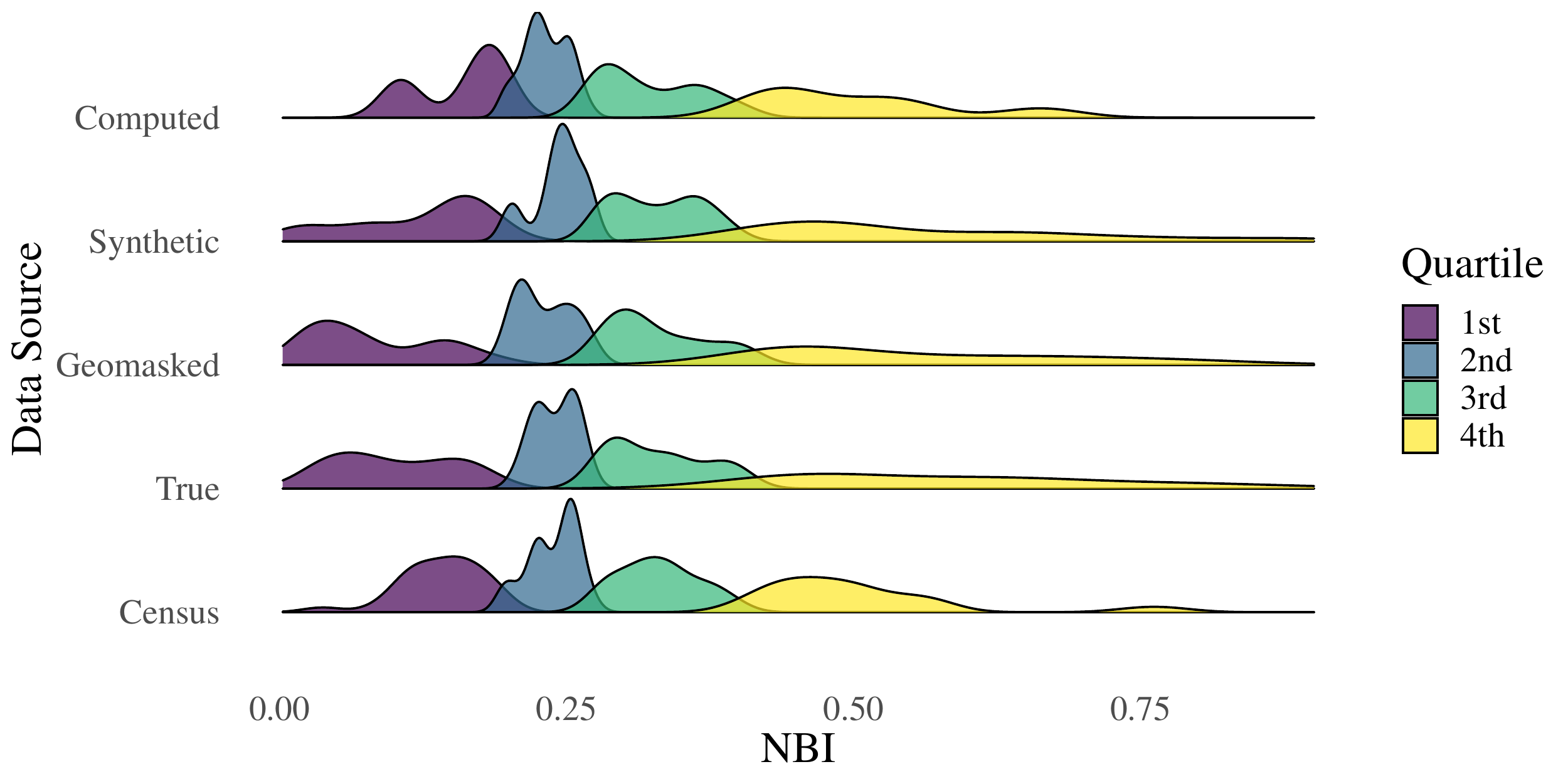}
         \caption{NBI densities}
         \label{fig:appendix-nbi-densities}
     \end{subfigure}
        \caption{\textbf{Performance of the synthetic vs. computed composite NBI.}\\ (a) - (c) show of the different survey types in our survey augmentation experiment across 100 simulation runs. (d) shows the densities of the composite NBI by quartiles for one simulation run.}
        \label{fig:appendix-nbi}
\end{figure}

Although the overall number of survey respondents with unsatisfied needs are captured with a high accuracy as measured by the normalized KL divergence $Z_{KL}$ for binary data, the NBI status on the individual level strongly diverges following Pearson's $\rho$ (cf. Supplementary Table \ref{table:appendix-pred}). Supplementary Figure \ref{fig:appendix-nbi} shows that the lack of linear correlation is mainly due to improperly captured relationships in the underlying variables than in the synthetic NBI as the former is outperformed by the latter for survey augmentation expressed in terms of adjusted $R^2$, bias and MSE. However, it remains on par with the geomasked survey at lower privacy risks.

\end{document}